\documentclass[aps,prx,preprint,onecolumn,superscriptaddress,nofootinbib,longbibliography]{revtex4-2}

\usepackage{amsmath,amssymb,bm,mathtools}
\usepackage{physics}
\usepackage{microtype}
\usepackage[hidelinks]{hyperref}
\usepackage{graphicx}

\newcommand{\Curl}{\operatorname{Curl}}

\newcommand{\Ael}{A^{\sigma}}

\begin{document}

\author{Alessio Zaccone}
\email{alessio.zaccone@unimi.it}
\affiliation{Department of Physics ``A.~Pontremoli'', University of Milan, via Celoria 16, 20133 Milan, Italy}

\author{Konrad Samwer}
\email{ksamwer@gwdg.de}
\affiliation{I.~Physikalisches Institut, University of G\"ottingen, Friedrich-Hund-Platz 1, 37077 G\"ottingen, Germany}

\title{Ginzburg-Landau Theory with Elastic Incompatibility Describes the Liquid--G-phase--Crystal State Diagram}

\date{\today}

\begin{abstract}
We derive a mechanically consistent Ginzburg--Landau theory for the
metastable heterogeneous G state reported in deeply undercooled metallic
liquids.  The central result is that type-II-like mixed-state physics can
emerge structurally with the random internal stress field playing the role of the magnetic field.
Frozen eigenstrain due to elastic incompatibility between ordered cores and the amorphous matrix generates incompatible stress, while the non-affine internal strain in the
liquid-like matrix in between the ordered cores relaxes that stress over a finite screening length
$\lambda_\sigma$ (analogous to the London length in the superconducting problem).  Structural order heals over a distinct mechanically-dressed length $\xi_G$; their ratio
$\kappa_G=\lambda_\sigma/\xi_G$ provides the structural counterpart of
the Ginzburg--Landau parameter.  Because the matrix is transiently rigid
yet locally relaxational only over an intermediate time window, the same
mechanism selects crystal for slow protocols, a heterogeneous G state at
intermediate rates, and a homogeneous L-like glass under rapid arrest.
Residual incompatibility frustrates coarsening, allowing finite ordered
cores and screened, orientation-dependent core interactions.  The
resulting texture is an inverted Abrikosov analogue: structural order resides in the
cores, whereas incompatible-stress screening is carried by the surrounding disordered matrix.
The theory yields direct simulation and experimental tests through
stress-decay, structural-healing, non-affine-susceptibility, and kinetic
measurements.
\end{abstract}

\maketitle

\section{Introduction}

Mixed states provide some of the clearest examples of how competing
length scales can reorganize a phase transition.  In a type-II
superconductor, the competition between the coherence length and the
magnetic penetration depth permits an intermediate vortex state rather
than a direct transition between spatially homogeneous phases.  A
natural question is whether an analogous mechanism can arise in
structural condensed matter problems, where the competing field is mechanical rather than
electromagnetic and where elastic compatibility and nonaffine displacements produce the screening mechanism.

Deeply undercooled metallic liquids offer a concrete setting in which
to ask this question. Molecular-dynamics studies of Ag, Cu--Ag, and
Cu--Zr report a first-order transformation from a homogeneous liquid
state (L) to a metastable, mechanically rigid and spatially
heterogeneous glassy state (G), in competition with crystallization into
the equilibrium crystal (X)
\cite{An2020,AnCuAg2020,An2021,AnCuZr2022}.  The G state is not a
homogeneous ``partly crystalline'' phase.  It consists of defective,
crystal-like ordered cores embedded in and separated by a continuous
disordered liquid-like matrix, while long-range translational order
remains absent. More broadly, frustration-based approaches to glass formation have emphasized
that locally preferred structural order may be prevented from extending
uniformly through space, leading to avoided ordering and glassy dynamics
\cite{Nussinov2004,TarjusKivelsonNussinovViot2005}. Simulations further show that formation of this state
is accompanied by a pronounced reduction of local deviatoric-strain
energy \cite{An2021}.

These observations pose a specific theoretical problem.  A random
internal stress cannot simply be inserted as the structural analogue of
a magnetic field: in elasticity the source is an incompatible
strain/distortion field, while the stress follows only after imposing a
constitutive law and mechanical equilibrium
\cite{Beltrami1886,Mura1987,Kroner1981}.  Moreover, ordinary
incompatible elasticity is long ranged and by itself does not generate
a London-like screening length.  A mechanically consistent structural
mixed state therefore requires three ingredients that must be kept
distinct: an incompatibility source, a stress field satisfying force
balance, and an additional relaxation channel capable of screening that
stress over a finite distance.

Here we show that these ingredients arise naturally by coupling a
first-order structural Ginzburg--Landau sector to incompatible elasticity
and a non-affine internal strain field.  Frozen or slowly evolving
eigenstrain produces a mechanically equilibrated incompatible stress,
while non-affine rearrangements in the liquid-like matrix reduce that
stress and generate a finite screening length $\lambda_\sigma$.
Structural order heals over a distinct, mechanically dressed length
$\xi_G$.  Their ratio $\kappa_G=\lambda_\sigma/\xi_G$ is the structural
counterpart of the superconducting Ginzburg--Landau parameter, but the
analogy is not imposed by replacing magnetic field with stress: both
lengths follow from the same coupled mechanical and structural
functional.

Several consequences follow from this construction.  First, the G state
occupies a kinetic window in which the surrounding material is rigid
enough on the observation time scale to store incompatibility energy but
still locally mobile enough to relax it non-affinely.  Second, the
ordered-core size is not identified with the healing length by hand;
finite cores emerge from the competition among structural driving,
interfacial cost, and residual incompatibility, while overlap of screened
residual stress fields can further frustrate coalescence.  Third, the
same framework produces a stress--temperature G-state wedge between
crystal and liquid, giving an inverted Abrikosov-like texture in which
order is concentrated in the cores whereas screening is carried by the
disordered matrix.  The theory therefore turns a superconducting analogy
into a falsifiable mechanical mechanism, with direct observables in
stress profiles, structural correlations, non-affine susceptibilities,
and relaxation times.

Throughout, the scalar field $\phi(\bm r)$ measures local structural
order: $\phi\simeq0$ denotes a liquid-like environment and
$\phi\simeq\phi_X$ an ordered/crystal-like environment.  The G state is
represented by a heterogeneous solution $\phi(\bm r)$ rather than by a
third spatially uniform value of the order parameter.  When $\phi_G$ is
used below, it denotes the local mechanically dressed ordered minimum
relevant to a G core.  We also distinguish this heterogeneous G state
from a rapidly quenched, comparatively homogeneous L-like glass.
\section{Central physical picture: a mechanically generated mixed state}

The type-II analogy becomes useful only after its mechanical content is made explicit.  In the superconducting mixed state, the order parameter and the screened field recover over two distinct lengths.  In the present structural problem there is no literal magnetic analogue of stress.  Instead, incompatible distortion generates a mechanically equilibrated stress, local non-affine rearrangements screen that stress in the disordered material, and structural order heals over a separate length.  The relevant question is therefore whether these two physical lengths, together with interfacial and incompatibility energetics, can select a metastable heterogeneous state with finite cores.

The central proposal is that the undercooled liquid contains geometrically incompatible local distortions.  If the liquid is fully relaxed on the observation time scale, viscous flow removes their long-lived elastic signature.  If it is completely frozen, stress can remain trapped but the local non-affine degrees of freedom can no longer reorganize into an equilibrated screening cloud.  The G state occupies the intermediate regime: the material has acquired a finite transient shear rigidity, so incompatibility carries a substantial elastic cost, while local rearrangements in the disordered matrix remain active and can reduce that cost around ordered regions.  The resulting state is the whole heterogeneous core--matrix texture: defective crystal-like cores coexist with and are separated by a structurally liquid-like, stress-accommodating matrix that need not flow on the observation time scale.

The correspondence with type-II superconductivity is therefore a correspondence between physical roles, not between microscopic fields:

\begin{align}
\text{normal phase} &\longleftrightarrow \text{liquid L},\\
\text{superconducting phase} &\longleftrightarrow \text{ordered/crystalline X},\\
\text{mixed phase} &\longleftrightarrow \text{heterogeneous G},\\
\xi &\longleftrightarrow \xi_G,\\
\lambda &\longleftrightarrow \lambda_\sigma.
\end{align}
Here $\xi_G$ is the structural healing scale (with finite cores expected to be of this order), whereas $\lambda_\sigma$ is the stress-screening scale. Throughout this paper, ``G phase'' refers to the complete heterogeneous state, not to the liquid-like matrix alone.

\subsection{The mixed state is an inverted Abrikosov analogue}

The analogy must be qualified at the level of the local core--shell morphology.  In an ordinary Abrikosov vortex the superconducting order parameter is suppressed in the vortex core and recovers outside it,
\begin{equation}
 |\psi|\simeq0\quad\text{in the core},
 \qquad
 |\psi|\to|\psi_{\rm SC}|\quad\text{outside},
\end{equation}
while magnetic induction is concentrated in the core and is screened by the superconducting exterior.  The structural texture considered here is the reverse in the order-parameter sector:
\begin{equation}
 \phi\simeq\phi_X\quad\text{in the ordered G core},
 \qquad
 \phi\to0\quad\text{in the liquid-like shell/exterior},
\end{equation}
and the non-affine screening response is strongest in the liquid-like material.  We therefore use the term \emph{inverted Abrikosov-like state}: the macroscopic dictionary L $\leftrightarrow$ normal, X $\leftrightarrow$ superconducting, and G $\leftrightarrow$ mixed state is retained, but the microscopic vortex morphology is inverted. More generally, the sharing of an order-parameter structure between
structural and superconducting systems does not imply identical defect
topology, since mechanical and geometrical constraints can fundamentally
alter the allowed defects \cite{KamienMosna2016}. Unless a quantized
Burgers/disclination content is demonstrated here, ``vortex'' should
therefore be understood only as an organizing analogy; the more literal
object is an ordered inclusion carrying or surrounded by incompatible
elastic content.

\section{From analogy to mechanism: incompatibility generates stress and relaxation screens it}

The superconducting analogy is used only to organize the two competing length scales.  In the London/GL problem the magnetic field is screened over $\lambda$, while the order parameter heals over $\xi$.  A direct structural transcription would be misleading, however, because physical stress is not the curl of an elastic gauge potential.  In continuum defect language the curl of an incompatible elastic distortion is a Nye/dislocation-density tensor, and the corresponding stress is obtained only after imposing the elastic constitutive law and mechanical equilibrium.

The detailed translational-gauge construction, together with the relation between the Nye tensor, strain incompatibility, and stress, is given in the Appendix.  The main result needed here is the hierarchy
\begin{equation}
 \begin{aligned}
 \text{incompatible distortion}
 &\longrightarrow \text{defect/incompatibility content}\\
 &\longrightarrow \text{mechanically equilibrated stress}.
 \end{aligned}
 \label{eq:incHierarchyMain}
\end{equation}
Ordinary incompatible elasticity is long ranged and does not by itself produce a London penetration depth.  A finite stress-screening length therefore requires an additional relaxational field.  We now introduce that field directly and use the resulting incompatible-elastic formulation as the main working theory.

\section{Incompatible-elastic Ginzburg--Landau theory with non-affine screening}

\subsection{Fields and free energy}

Let $\phi(\bm r)$ be a scalar amplitude of local structural order,
\begin{equation}
\phi\simeq0\quad \text{liquid-like},
\qquad
\phi\simeq\phi_X>0\quad \text{ordered/crystal-like}.
\end{equation}
No assumption about fcc, hcp, icosahedral, or other detailed symmetry is required at this level.

Let $\epsilon^*_{ij}(\bm r)$ be a frozen or slowly evolving eigenstrain encoding the random structural mismatch. Let $p_{ij}(\bm r)$ be a non-affine relaxational strain that can develop much more readily in the liquid-like shell than in an ordered core. Non-affine displacements and the associated non-affine elastic response are generic features of disordered and amorphous solids, where local particle motions deviate from the affine deformation imposed at macroscopic scales \cite{Alexander,DiDonnaLubensky2005,Lemaitre,ZacconeScossaRomano2011,ZacconeBook2023}. Define
\begin{equation}
 e_{ij}
 =\frac{1}{2}(\partial_i u_j+\partial_j u_i)
 -\epsilon^*_{ij}
 -p_{ij}.
\label{eq:etotal}
\end{equation}
Taking the incompatibility Saint-Venant condition \cite{Beltrami1886,Mura1987,Kroner1981,Zaccone_book_2026} gives
\begin{equation}
\bm\nabla\times\bigl(\bm\nabla\times e\bigr)
=-\bm\nabla\times\Bigl[\bm\nabla\times\bigl(\epsilon^*+p\bigr)\Bigr]
\equiv \eta^0+\eta^p,
\label{eq:screenedinc}
\end{equation}
where
\begin{equation}
\eta^0=-\bm\nabla\times\bigl(\bm\nabla\times\epsilon^*\bigr),
\qquad
\eta^p=-\bm\nabla\times\bigl(\bm\nabla\times p\bigr).
\end{equation}
Equation~\eqref{eq:screenedinc} is the central structural equation: the frozen incompatibility is not simply removed; it is compensated by a relaxational incompatibility carried by non-affine rearrangements.

The minimal free energy is
\begin{align}
F_{II}[\phi,\bm u,p;\epsilon^*]
=\int\dd^3r\,\Bigg\{&
 f_0(\phi)
 +\frac{K_\phi}{2}|\nabla\phi|^2
 +\frac{1}{2}C_{ijkl}(\phi)e_{ij}e_{kl}
\nonumber\\
&+\frac{A(\phi)}{2}p^D_{ij}p^D_{ij}
+\frac{D(\phi)}{2}\partial_kp^D_{ij}\partial_kp^D_{ij}
\Bigg\}.
\label{eq:FormII}
\end{align}
Here $p^D$ is the deviatoric part of $p$. The coefficients have direct interpretations:
 $C_{ijkl}$ are the elastic moduli; $A$: is the local restoring stiffness against non-affine deviatoric rearrangement;
 $D$ is the gradient stiffness of the non-affine strain field, with units of modulus times length squared;
$K_\phi$ is the usual structural-order gradient stiffness of structural GL theory \cite{Cowley,ChaikinLubensky1995}.

The simplest physical assumption is
\begin{equation}
A_L\ll A_X,
\end{equation}
so that non-affine relaxation is easy in the liquid-like shell and strongly suppressed in the ordered core.

Variation with respect to $u_i$ gives mechanical equilibrium,
\begin{equation}
\partial_j\sigma_{ij}=0,
\qquad
\sigma_{ij}=C_{ijkl}(\phi)e_{kl}.
\label{eq:mecheqII}
\end{equation}
Thus compatibility/incompatibility and force balance are retained exactly even though the detailed gauge field has been eliminated.

\section{Mechanical projection, a precise deviatoric channel, and stress screening}

\subsection{Mechanical projection of the compatible displacement}

The compatible displacement $u_i$ should not be counted as an independent source of elastic frustration.  Define
\begin{equation}
 q_{ij}\equiv\epsilon^*_{ij}+p_{ij},
 \qquad
 e_{ij}=\epsilon_{ij}(u)-q_{ij}.
\end{equation}
For fixed $q_{ij}$, minimizing the elastic energy with respect to $u_i$ enforces mechanical equilibrium, $\partial_j\sigma_{ij}=0$, and removes the part of $q_{ij}$ that can be represented by a single-valued displacement.  Symbolically one obtains
\begin{equation}
 F_{\rm el}^{\rm proj}
 =\frac12\int_{\bm k}q^\dagger
 \mathcal K(\bm k;\phi)q,
 \label{eq:FelK}
\end{equation}
where $\mathcal K$ is the mechanically projected elastic kernel.  Its defining property is
\begin{equation}
 q\ \text{compatible}\quad\Longrightarrow\quad\mathcal Kq=0.
\end{equation}
Thus the projection does not eliminate incompatibility; it does the opposite.  It removes the elastically redundant compatible part and leaves precisely the content that cannot be generated by a smooth displacement field and therefore produces long-ranged internal stress.  In the following we do not need the explicit matrix form of $\mathcal K$; we only need a mechanically admissible deviatoric eigenchannel of this projected problem.

\subsection{A mathematically definite one-channel projection}

The deviatoric strain space in three dimensions is five dimensional.  A scalar reduction is meaningful only after a definite tensor direction in this space has been specified.  For a planar interface with unit normal $\bm n$, choose two orthonormal tangent vectors $\bm t_1$ and $\bm t_2$ and define
\begin{equation}
 Q_{ij}=\frac12
 \left(t_{1i}t_{1j}-t_{2i}t_{2j}\right).
 \label{eq:Qdef}
\end{equation}
Then
\begin{equation}
 Q_{ii}=0,
 \qquad
 Q_{ij}n_j=0,
 \qquad
 Q_{ij}Q_{ij}=\frac12.
 \label{eq:Qprops}
\end{equation}
The condition $Q\bm n=0$ is important: a stress field proportional to $Q_{ij}$ and depending only on the coordinate normal to the interface automatically satisfies force balance.  For $\bm n=\hat{\bm x}$ one convenient choice is
\begin{equation}
 Q=\frac12
 \begin{pmatrix}
 0&0&0\\
 0&1&0\\
 0&0&-1
 \end{pmatrix}.
\end{equation}

For any deviatoric tensor $X^D_{ij}$, the amplitude in this channel is the coefficient of $Q$,
\begin{equation}
 X_Q=\frac{Q_{ij}X^D_{ij}}{Q_{mn}Q_{mn}}
 =2Q_{ij}X^D_{ij}.
 \label{eq:Qprojection}
\end{equation}
The conjugate projected stress is $\sigma_Q=Q_{ij}\sigma^D_{ij}$.  In an isotropic material, if $e^D_{ij}=e_Q Q_{ij}$, then
\begin{equation}
 f_{\rm el}=G\,e^D:e^D
 =\frac{G}{2}e_Q^2,
 \qquad
 \sigma_Q=G e_Q,
\end{equation}
so the coefficient denoted below by $\mu_L$ is the ordinary shear modulus $G_L$ with the normalization in Eq.~\eqref{eq:Qdef}.

More generally, in an anisotropic system one first performs the mechanical projection summarized in Eq.~\eqref{eq:FelK}, restricts the resulting kernel to the five-dimensional deviatoric subspace, and diagonalizes it.  Choosing one eigenvector $v_\alpha(\hat{\bm k})$ defines
\begin{equation}
 Q_{ij}=\sum_{\alpha=1}^{5}v_\alpha Q^{(\alpha)}_{ij}.
\end{equation}
The scalar model is then the one-eigenmode truncation of the full mechanically projected tensor theory, rather than an unspecified ``deviatoric component''.

\subsection{Reduced incompatible-elastic channel}

Let $e_0(x)$ be the amplitude in the chosen channel of the elastic strain generated by the frozen incompatibility after the compatible displacement has been eliminated.  Choose the sign of the scalar relaxational amplitude $p(x)$ so that positive $p$ cancels positive $e_0$.  In terms of the tensor field used in Eq.~\eqref{eq:etotal}, this amounts to writing the projected relaxational eigenstrain as $p^D_{ij}=-pQ_{ij}$.  The residual elastic strain in the channel is then $(e_0-p)Q_{ij}$.

In a locally liquid-like region the projected stress sector is
\begin{equation}
 F_{\sigma}^{\rm red}
 =\int\dd x\left[
 \frac{\mu_L}{2}(e_0-p)^2
 +\frac{A_L}{2}p^2
 +\frac{D_L}{2}(p')^2
 \right].
 \label{eq:reducedstress}
\end{equation}
The generalized scalar stress conjugate to $e_0-p$ is
\begin{equation}
 \sigma=\mu_L(e_0-p),
 \qquad
 h_\sigma\equiv\mu_L e_0.
 \label{eq:sigmah}
\end{equation}
Here $h_\sigma$ is not an arbitrary random stress: it is the mechanically projected bare stress generated by the frozen incompatibility $\eta^0$ before the non-affine mode is allowed to relax.

Variation with respect to $p$ gives
\begin{equation}
 -D_Lp''+(\mu_L+A_L)p=h_\sigma.
 \label{eq:p1d}
\end{equation}
Hence
\begin{equation}
 \lambda_\sigma^2=\frac{D_L}{\mu_L+A_L}.
 \label{eq:lambdadef}
\end{equation}
In a source-free liquid shell, $h_\sigma=0$, both $p$ and the projected stress obey a Helmholtz equation and decay exponentially over $\lambda_\sigma$.

\subsection{How much incompatible strain is accommodated? The distinct roles of \texorpdfstring{$\mu_L$ and $A_L$}{muL and AL}}

Before discussing the spatial decay, it is useful to separate the \emph{amount} of local accommodation from the \emph{distance} over which that accommodation is cooperative.  Consider first a spatially uniform projected incompatible strain $e_0$ and neglect the gradient term.  The local free-energy density is
\begin{equation}
 f_{\sigma}^{(0)}(p)
 =\frac{\mu_L}{2}(e_0-p)^2+\frac{A_L}{2}p^2.
 \label{eq:localpenergy}
\end{equation}
Stationarity with respect to $p$ gives
\begin{equation}
 -\mu_L(e_0-p)+A_Lp=0,
\end{equation}
and hence
\begin{equation}
 p_0=\frac{\mu_L}{\mu_L+A_L}e_0.
 \label{eq:p0uniform}
\end{equation}
This motivates the dimensionless \emph{non-affine accommodation fraction}
\begin{equation}
 \mathcal S_L\equiv\frac{p_0}{e_0}
 =\frac{\mu_L}{\mu_L+A_L},
 \qquad 0\leq\mathcal S_L\leq1.
 \label{eq:screenfraction}
\end{equation}
The residual elastic strain and stress are then
\begin{equation}
 e_{\rm res}=e_0-p_0
 =\frac{A_L}{\mu_L+A_L}e_0,
 \label{eq:eres}
\end{equation}
\begin{equation}
 \sigma_{\rm res}
 =\mu_L e_{\rm res}
 =\frac{\mu_LA_L}{\mu_L+A_L}e_0
 =h_\sigma\frac{A_L}{\mu_L+A_L}.
 \label{eq:sigmares}
\end{equation}
Thus
\begin{equation}
 \frac{\sigma_{\rm res}}{h_\sigma}
 =1-\mathcal S_L
 =\frac{A_L}{\mu_L+A_L}.
 \label{eq:stressfraction}
\end{equation}
The distinction between the absolute stress and the stress \emph{relative to the bare incompatible stress} is important.  At fixed $e_0$, increasing $\mu_L$ increases the bare stress $h_\sigma=\mu_Le_0$.  The fraction of that stress left unscreened decreases, but the absolute residual stress in Eq.~\eqref{eq:sigmares} need not decrease; it approaches $A_Le_0$ for $\mu_L\gg A_L$.  Therefore a larger transient shear modulus means a stronger energetic drive to accommodate incompatible strain, not necessarily a smaller absolute stress amplitude.

The same result is visible energetically.  Without the non-affine response,
\begin{equation}
 f_{\rm unrel}=\frac{\mu_L}{2}e_0^2,
\end{equation}
whereas after minimization,
\begin{equation}
 f_{\rm rel}
 =\frac{\mu_LA_L}{2(\mu_L+A_L)}e_0^2.
\end{equation}
The energy density released by non-affine accommodation is therefore
\begin{equation}
 \Delta f_{\rm scr}
 =f_{\rm unrel}-f_{\rm rel}
 =\frac{\mu_L^2e_0^2}{2(\mu_L+A_L)}
 =\frac12\mu_Le_0^2\mathcal S_L.
 \label{eq:deltafscrStatic}
\end{equation}
Consequently $A_L$ and $\mu_L$ have different meanings.  The modulus $\mu_L$ measures the energetic cost of leaving the incompatible strain stored elastically.  The coefficient $A_L$ is the local curvature of the free energy in the non-affine coordinate $p$; equivalently, it is an inverse local non-affine susceptibility.  A small $A_L$ permits a large equilibrium non-affine response.  This static energetic stiffness should not be confused with the kinetic mobility of that response, which is introduced below.

\subsection{Transient liquid rigidity, local mobility, and the external time scale}

An equilibrium liquid has zero static shear modulus.  Therefore $\mu_L$ in the present theory cannot be interpreted as $G(\omega=0)$.  It is the \emph{recoverable} shear rigidity seen on the externally available observation or transformation time $\tau_{\rm obs}$.  Operationally,
\begin{equation}
 \mu_L(T,\dot T)\simeq G'(\omega_{\rm obs},T),
 \qquad
 \omega_{\rm obs}\sim\tau_{\rm obs}^{-1},
 \label{eq:muStorage}
\end{equation}
or, equivalently, it may be taken from the plateau of a shear-stress relaxation function on that time scale.  The use of $G'$ rather than $G''$ is deliberate: $G'$ measures the recoverable energy stored elastically and is therefore the quantity that enters the static free-energy term $\mu_L(e_0-p)^2/2$, whereas $G''$ measures dissipative loss.  Near a viscoelastic crossover both can be comparable and both are experimentally informative, but only the recoverable part should be identified with the elastic coefficient $\mu_L$.

For a single Maxwell element,
\begin{align}
 G'(\omega,T)
 &=G_\infty(T)\frac{[\omega\tau_\alpha(T)]^2}
 {1+[\omega\tau_\alpha(T)]^2},\\
 G''(\omega,T)
 &=G_\infty(T)\frac{\omega\tau_\alpha(T)}
 {1+[\omega\tau_\alpha(T)]^2},
 \label{eq:MaxwellStorage}
\end{align}
with $\tau_\alpha$ the bulk structural $\alpha$-relaxation time \cite{Ferry1980}.  These expressions are used only as the simplest illustration of the time-scale dependence; the theory does not require a single-Maxwell liquid.

The non-affine field has its own coarse-grained kinetics. A long-time
relaxational closure of time-dependent Ginzburg--Landau form, consistent
with the static free energy \cite{Onuki2002}, is
\begin{equation}
 \partial_t p
 =-\Gamma_p\frac{\delta F}{\delta p}
 =\Gamma_p
 \left[D_L\nabla^2p-(\mu_L+A_L)p+h_\sigma\right].
  \label{eq:pdynamics}
\end{equation}
This equation is not intended to describe the ballistic or vibrational regime of atomic motion.  It is an overdamped coarse-grained description after those fast microscopic motions have been averaged out.  For the uniform mode it gives the local response time
\begin{equation}
 \tau_p=\frac{1}{\Gamma_p(\mu_L+A_L)}.
 \label{eq:taup}
\end{equation}
Physically, $\tau_p$ is the relaxation time of the particular local non-affine channel that screens the projected incompatibility.  It need not be identical to the bulk $\tau_\alpha$, but it is natural to expect it to be comparable to a \emph{local} structural/non-affine relaxation time $\tau_{\rm loc}$ of the liquid-like matrix.  We do not identify $\tau_{\rm loc}$ a priori with the Johari--Goldstein $\beta$-relaxation time $\tau_\beta$ \cite{JohariGoldstein1970}.  A Johari--Goldstein-like secondary relaxation is a plausible microscopic realization of such a faster local process. We therefore write
\begin{equation}
 \tau_p\sim\tau_{\rm loc},
 \qquad
 \tau_{\rm loc}\lesssim\tau_\alpha^{\rm bulk}
 \quad\text{in the G-forming regime}.
 \label{eq:tauplocal}
\end{equation}
This hierarchy should be measured rather than assumed; in particular, comparison of $\tau_p$ with any independently identified $\tau_\beta$ provides a direct test of whether the Johari--Goldstein process participates in the local stress-accommodation dynamics.

In Fourier-frequency space, Eq.~\eqref{eq:pdynamics} gives
\begin{equation}
 p(\bm k,\omega)
 =\frac{h_\sigma(\bm k,\omega)}
 {\mu_L+A_L+D_Lk^2-i\omega/\Gamma_p}.
 \label{eq:pdynamicSusceptibility}
\end{equation}
The static screening length therefore applies when the local non-affine mode can follow the externally imposed evolution, approximately $\omega_{\rm obs}\tau_p\ll1$.

The experimentally or computationally controlled quantity is $\tau_{\rm obs}$: the time available to the system in the relevant temperature interval.  Under continuous cooling one may estimate
\begin{equation}
 \tau_{\rm obs}\sim\frac{\Delta T_G}{|\dot T|},
 \label{eq:tauCooling}
\end{equation}
whereas in an isothermal MD protocol $\tau_{\rm obs}$ is simply the hold/observation time.  The G-phase window is then most transparently written as
\begin{equation}
 \tau_p\sim\tau_{\rm loc}
 \lesssim\tau_{\rm obs}
 \lesssim\tau_\alpha^{\rm bulk},
 \label{eq:Gtimewindow}
\end{equation}
together with the requirement that complete crystallization not outrun the protocol.  If $\tau_X$ denotes the relevant crystal nucleation-and-growth time, the latter condition is schematically $\tau_X\gtrsim\tau_{\rm obs}$ during G formation.

The meaning of Eq.~\eqref{eq:Gtimewindow} is simple.  The right inequality means that the bulk liquid cannot globally flow and erase the incompatible elastic field during the observation time, so it possesses a finite transient rigidity.  The left inequality means that local rearrangements in the liquid-like matrix are still fast enough to accommodate part of that field.  Thus the matrix is \emph{structurally liquid-like but mechanically viscoelastic}: it does not flow globally on $\tau_{\rm obs}$, yet it can still rearrange locally.

Two limiting cases are useful.  For sufficiently long observation times, $\tau_\alpha^{\rm bulk}\lesssim\tau_{\rm obs}$, global relaxation is accessible; if also $\tau_X\lesssim\tau_{\rm obs}$, classical crystal nucleation and growth can proceed and the X phase is reached.  At the opposite extreme, $\tau_{\rm obs}\lesssim\tau_p$ while $\tau_\alpha^{\rm bulk}\gg\tau_{\rm obs}$, even the local screening response cannot equilibrate.  Rapid quenching then freezes a comparatively homogeneous \emph{L-like glass}, distinct from the heterogeneous G phase.

Increasing the cooling rate shortens $\tau_{\rm obs}$.  This can increase the apparent recoverable rigidity because the material is probed at a higher effective frequency, while simultaneously reducing the time available for $p$ to respond.  These opposing effects make a finite cooling-rate/observation-time window for G-phase formation a natural possibility of the theory.

\subsection{Non-affine cooperativity length and the role of chemistry}

The formal screening length
\begin{equation}
 \lambda_\sigma=\sqrt{\frac{D_L}{\mu_L+A_L}}
\end{equation}
can be misleading if one varies $\mu_L$ while silently holding $D_L$ fixed.  It is more transparent to define, whenever $\mu_L>0$, the non-affine cooperativity length
\begin{equation}
 \ell_{\rm NA}^2\equiv\frac{D_L}{\mu_L}.
 \label{eq:ellNAdef}
\end{equation}
This is simply a reparameterization of the gradient stiffness and does not require an additional approximation.  Using Eq.~\eqref{eq:screenfraction},
\begin{equation}
 \lambda_\sigma
 =\ell_{\rm NA}
 \sqrt{\frac{\mu_L}{\mu_L+A_L}}
 =\ell_{\rm NA}\sqrt{\mathcal S_L}.
 \label{eq:lambdaEllNA}
\end{equation}
Equation~\eqref{eq:lambdaEllNA} separates two physically different aspects of screening.  The factor $\mathcal S_L$ measures \emph{how much} of a uniform incompatible strain can be accommodated, whereas $\ell_{\rm NA}$ measures \emph{how far} the non-affine accommodation is spatially cooperative.

At fixed $D_L$ and $A_L$, increasing $\mu_L$ indeed decreases $\lambda_\sigma$.  However, this partial derivative is not the appropriate statement when comparing different liquid chemistries, because $D_L$ will generally change together with $\mu_L$.  If chemistry changes $\mu_L$ while leaving $\ell_{\rm NA}$ and $A_L$ approximately fixed, then $\mathcal S_L$ increases and $\lambda_\sigma$ approaches $\ell_{\rm NA}$ from below.  If instead $D_L$ remains approximately fixed, then $\ell_{\rm NA}\propto\mu_L^{-1/2}$ and the shell becomes thinner.  Thus there is no universal monotonic dependence of the shell thickness on $\mu_L$ alone.

Changing chemistry can modify at least four quantities relevant to the screening mechanism:
\begin{equation}
 \mu_L,\qquad A_L,\qquad \ell_{\rm NA},\qquad e_0,
\end{equation}
and it may also modify the structural parameters that determine the dressed core scale $\xi_G$.  The first two determine the static accommodation fraction $\mathcal S_L$; $\ell_{\rm NA}$ determines the spatial cooperativity of that accommodation; and $e_0$ sets the strength of the incompatible source through $h_\sigma=\mu_Le_0$.  At fixed $e_0$, the energy available from stress relaxation is Eq.~\eqref{eq:deltafscrStatic}, which can increase substantially with $\mu_L$ even when the spatial screening length does not.  Consequently the propensity to form a G texture cannot be inferred from the liquid shear modulus alone.

Combining Eqs.~\eqref{eq:kappa_def} and \eqref{eq:lambdaEllNA} gives the useful representation
\begin{equation}
 \kappa_G
 =\frac{\ell_{\rm NA}}{\xi_G}
 \sqrt{\mathcal S_L}
 =\frac{\ell_{\rm NA}}{\xi_G}
 \sqrt{\frac{\mu_L}{\mu_L+A_L}}.
 \label{eq:kappaEllNA}
\end{equation}
This form makes the material interpretation especially clear: a large $\kappa_G$ requires a non-affine cooperative length large compared with the structural core scale and an appreciable ability of the viscoelastic liquid to accommodate incompatible strain.  Cooling rate determines whether this static ability is dynamically accessible through Eq.~\eqref{eq:Gtimewindow}.

\section{First-order structural GL sector and the coherence length}

Elastic coupling is well known to modify the character of structural
transitions and can, under appropriate conditions, cause an otherwise
continuous transition to be first order \cite{LarkinPikin1969}.  Landau theory has also been applied directly to first-order
liquid--crystal ordering and crystal nucleation
\cite{AlexanderMcTague}. Here we do not attempt to derive either
mechanism microscopically; instead, we encode the observed first-order
liquid--ordered competition phenomenologically through the minimal
sixth-order structural potential,
\begin{equation}
 f_0(\phi)
=\frac{a(T)}{2}\phi^2
-\frac{b}{4}\phi^4
+\frac{c}{6}\phi^6,
\qquad b,c>0.
\label{eq:phi6}
\end{equation}
The structural free energy is
\begin{equation}
F_\phi=\int\dd^3r\left[
\frac{K_\phi}{2}|\nabla\phi|^2+f_0(\phi)
\right].
\end{equation}
The stationary uniform states satisfy
\begin{equation}
\frac{\partial f_0}{\partial\phi}
=\phi(a-b\phi^2+c\phi^4)=0.
\label{eq:stationaryphi}
\end{equation}
For the ordered branch define $x\equiv\phi^2$. Then
\begin{equation}
 c x^2-bx+a=0,
\end{equation}
so
\begin{equation}
\phi_{\pm}^2
=\frac{b\pm\sqrt{b^2-4ac}}{2c}.
\label{eq:phibranches}
\end{equation}
The first-order coexistence point is determined by the simultaneous conditions
\begin{equation}
 f_0(\phi_X)=f_0(0)=0,
\qquad
 f_0'(\phi_X)=0.
\end{equation}
Using $x_X=\phi_X^2$, the stationary condition gives
\begin{equation}
 a=bx_X-cx_X^2.
\end{equation}
Substitution into $f_0=0$ gives
\begin{equation}
0=\frac{b}{4}x_X^2-\frac{c}{3}x_X^3,
\end{equation}
and therefore
\begin{equation}
\phi_X^2=\frac{3b}{4c},
\qquad
 a_{\rm coex}=\frac{3b^2}{16c}.
\label{eq:coex}
\end{equation}
At coexistence the potential factorizes exactly,
\begin{equation}
 f_0(\phi)
=\frac{c}{6}\phi^2(\phi^2-\phi_X^2)^2.
\label{eq:factorized}
\end{equation}

\subsection{From the bare crystal--liquid landscape to the dressed G-state coherence length}

It is useful to distinguish two energy landscapes. The function $f_0(\phi)$ describes the \emph{intrinsic structural} tendency of the material in the absence of the incompatibility-induced mechanical bias. It is the Landau landscape for the underlying liquid--ordered transition: at high temperature the liquid minimum near $\phi=0$ is favored, while on cooling the ordered minimum near $\phi=\phi_X$ becomes competitive and ultimately favorable. The coefficients $a(T)$, $b$, and $c$ characterize this structural competition only.

For the bare structural functional
\begin{equation}
 F_\phi[\phi]
 =\int\dd^3r\left[
 \frac{K_\phi}{2}|\nabla\phi|^2+f_0(\phi)
 \right],
 \label{eq:FphiBare}
\end{equation}
a variation $\phi\rightarrow\phi+\delta\phi$ gives
\begin{align}
 \delta F_\phi
 &=\int\dd^3r\left[
 K_\phi\nabla\phi\cdot\nabla\delta\phi
 +f_0'(\phi)\delta\phi
 \right]\\
 &=\int\dd^3r\left[
 -K_\phi\nabla^2\phi+f_0'(\phi)
 \right]\delta\phi,
\end{align}
where the surface term has been taken to vanish. The bare structural Euler--Lagrange equation is therefore
\begin{equation}
 -K_\phi\nabla^2\phi+f_0'(\phi)=0.
 \label{eq:phiGLEbare}
\end{equation}
For the sixth-order potential in Eq.~\eqref{eq:phi6},
\begin{equation}
 -K_\phi\nabla^2\phi+a\phi-b\phi^3+c\phi^5=0.
 \label{eq:phi6GLE}
\end{equation}

The G phase, however, does not live in this stress-free landscape. Its local structural environment is coupled to an incompatible elastic source. The simplest way to see the consequence without introducing additional formal machinery is to consider the local part of the projected mechanical energy,
\begin{equation}
 f_{\rm mech}(\phi,p)
 =\frac{\mu_L}{2}(e_0-p)^2+\frac{A(\phi)}{2}p^2,
 \label{eq:localMechPhi}
\end{equation}
where the local non-affine stiffness is smaller in liquid-like material than in ordered material,
\begin{equation}
 A(0)=A_L,\qquad A(\phi_X)=A_X,\qquad A_X\gg A_L.
\end{equation}
For fixed local incompatible strain $e_0$, minimization with respect to $p$ gives
\begin{equation}
 p_*(\phi)=\frac{\mu_L e_0}{\mu_L+A(\phi)}
 =\frac{h_\sigma}{\mu_L+A(\phi)},
 \qquad h_\sigma\equiv\mu_L e_0.
 \label{eq:pLocalPhi}
\end{equation}
Substituting this locally relaxed value back into the energy gives, up to a $\phi$-independent constant, the mechanically dressed structural landscape
\begin{equation}
 f_G(\phi;h_\sigma)
 =f_0(\phi)
 -\frac{h_\sigma^2}{2[\mu_L+A(\phi)]}.
 \label{eq:fGlocal}
\end{equation}
Equation~\eqref{eq:fGlocal} is not meant to replace the full spatially coupled theory; it is the simplest local representation of how incompatible stress reshapes the structural energy landscape. Its physical meaning is transparent. The bare function $f_0$ asks which structure would be preferred if mechanical frustration were absent. The second term asks how much of the incompatibility energy can be released by local non-affine accommodation. Because $A_L<A_X$, this relaxation term lowers the energy of liquid-like configurations more strongly than that of crystal-like configurations. Incompatible stress therefore competes with the bare drive to crystallize: it can raise the relative cost of a uniformly ordered state and make it favorable to retain liquid-like material around an ordered region. This is the local energetic origin of the G core--shell texture in the present theory.

The dressed G-state minimum $\phi_G$ is defined by
\begin{equation}
 f_G'(\phi_G)=0.
 \label{eq:phiGmin}
\end{equation}
Here $\phi_G$ is the \emph{local} dressed ordered value relevant to a G core.  It should not be interpreted as a homogeneous intermediate order parameter for the entire G phase, which is represented by the spatially heterogeneous core--matrix texture.
The curvature of this minimum measures how strongly the local state resists a structural fluctuation away from $\phi_G$. The physical structural healing length is therefore
\begin{equation}
 \xi_G
 =\sqrt{\frac{K_\phi}{f_G''(\phi_G)}}.
 \label{eq:xi_general}
\end{equation}
Thus $\xi_G$ is not fixed by the bare crystal--liquid potential alone. Through $f_G''$, it can depend on the magnitude of the incompatible stress, on the transient liquid modulus $\mu_L$, and on the way the non-affine stiffness $A(\phi)$ changes between liquid-like and ordered environments. Differentiating Eq.~\eqref{eq:fGlocal} makes this dependence explicit,
\begin{equation}
 f_G''(\phi)
 =f_0''(\phi)
 +\frac{h_\sigma^2}{2}
 \left[
 \frac{A''(\phi)}{[\mu_L+A(\phi)]^2}
 -\frac{2[A'(\phi)]^2}{[\mu_L+A(\phi)]^3}
 \right].
 \label{eq:fGcurvature}
\end{equation}
The sign of the correction is not universal because it depends on the detailed interpolation $A(\phi)$, but the important qualitative point is universal within the model: incompatible stress changes both the location and the curvature of the G-state minimum. Consequently the core size can change with cooling rate and chemistry even when the bare structural coefficients are unchanged, because $\mu_L$ and $h_\sigma$ are themselves time-scale- and material-dependent quantities.

For reference, in the stress-free limit $h_\sigma\rightarrow0$ one has $f_G\rightarrow f_0$, $\phi_G\rightarrow\phi_X$, and the dressed length reduces to the bare structural length
\begin{equation}
 \xi_0
 \equiv\sqrt{\frac{K_\phi}{f_0''(\phi_X)}}
 =\sqrt{\frac{4K_\phi c}{3b^2}}
 \qquad\text{at coexistence}.
 \label{eq:xi0_coex}
\end{equation}
We use $\xi_0$ as an analytic reference and reserve $\xi_G$ for the actual healing/core scale in the mechanically dressed G state.

The full coupled $\phi$ equation remains the variation of Eq.~\eqref{eq:FormII},
\begin{equation}
 \begin{aligned}
 -K_\phi\nabla^2\phi+f_0'(\phi)
 &+\frac12\frac{\partial C_{ijkl}}{\partial\phi}e_{ij}e_{kl}
 +\frac12A'(\phi)p^D_{ij}p^D_{ij}\\
 &+\frac12D'(\phi)\partial_kp^D_{ij}\partial_kp^D_{ij}=0.
 \end{aligned}
 \label{eq:phiGLEfull}
\end{equation}
Equation~\eqref{eq:fGlocal} should therefore be read as an intuitive local dressing of $f_0$, while Eq.~\eqref{eq:phiGLEfull} is the more general spatial statement.

\section{Structural GL ratio}

The structural analogue of the superconducting GL ratio is defined from the two \emph{physical} lengths,
\begin{equation}
 \kappa_G\equiv\frac{\lambda_\sigma}{\xi_G}.
 \label{eq:kappa_def}
\end{equation}
Using the local dressed curvature in Eq.~\eqref{eq:xi_general} gives
\begin{equation}
 \kappa_G
 =\sqrt{
 \frac{D_L}{\mu_L+A_L}
 \frac{f_G''(\phi_G)}{K_\phi}
 }.
 \label{eq:kappa_general}
\end{equation}
Equivalently, introducing $\ell_{\rm NA}^2=D_L/\mu_L$ and the accommodation fraction $\mathcal S_L=\mu_L/(\mu_L+A_L)$,
\begin{equation}
 \kappa_G
 =\frac{\ell_{\rm NA}}{\xi_G}
 \sqrt{\mathcal S_L}
 =\frac{\ell_{\rm NA}}{\xi_G}
 \sqrt{\frac{\mu_L}{\mu_L+A_L}}.
 \label{eq:kappaDressed}
\end{equation}
The stress-free structural result is recovered only as a reference limit: when $h_\sigma\to0$, $\xi_G\to\xi_0$ and Eq.~\eqref{eq:kappa_general} reduces to the expression obtained from $f_0''(\phi_X)$. In the actual G state, however, both the screening length and the structural core length can respond to cooling rate and chemistry. Equation~\eqref{eq:kappa_def}, based on directly measured $\lambda_\sigma$ and $\xi_G$, is therefore more fundamental than any formula in terms of coarse-grained coefficients.

Small $\kappa_G$ means that incompatible stress is accommodated over a distance shorter than the structural healing scale. Large $\kappa_G$ means that the stress-accommodation cloud extends substantially beyond the ordered-core scale. The analogy remains an \emph{inverted} Abrikosov one: structural order is concentrated in the core, while the liquid-like exterior is the principal screening medium.

\section{Physical mechanism: the G phase between crystal and the L-like glass}
\label{sec:mechanism}

The preceding equations can be summarized in terms of three intrinsic/external clocks and one competing transformation time.  The bulk liquid relaxes on $\tau_\alpha^{\rm bulk}$; the liquid-like matrix has a more local non-affine response time $\tau_p\sim\tau_{\rm loc}$; the cooling or holding protocol supplies the externally available time $\tau_{\rm obs}$; and crystal nucleation plus growth requires a time $\tau_X$.  The experimentally relevant control variable is therefore $\tau_{\rm obs}$ rather than a new intrinsic ``G clock''.

When $\tau_\alpha^{\rm bulk}\lesssim\tau_{\rm obs}$, the liquid has enough time to relax on large scales.  Incompatibility-generated stresses can be relaxed by viscous and structural rearrangement.  If crystallization is thermodynamically favored and $\tau_X\lesssim\tau_{\rm obs}$, ordered regions can undergo ordinary nucleation and growth and the limiting state is the crystal X.  This is the slow-cooling or long-observation-time side of the problem.

The response is qualitatively different in the window
\begin{equation}
 \tau_p\sim\tau_{\rm loc}
 \lesssim\tau_{\rm obs}
 \lesssim\tau_\alpha^{\rm bulk},
 \label{eq:GwindowSummary}
\end{equation}
provided crystallization is slower than the available time, $\tau_X\gtrsim\tau_{\rm obs}$.  Because $\tau_{\rm obs}\lesssim\tau_\alpha^{\rm bulk}$, the surrounding matrix appears elastically rigid on the protocol time scale and possesses a finite recoverable modulus $\mu_L\simeq G'(\omega_{\rm obs},T)$.  Frozen incompatibility therefore carries a real elastic-energy penalty, with bare projected stress $h_\sigma=\mu_Le_0$.  At the same time $\tau_p\lesssim\tau_{\rm obs}$, so local non-affine rearrangements can still occur.  They accommodate the fraction
\begin{equation}
 \mathcal S_L=\frac{\mu_L}{\mu_L+A_L}
\end{equation}
of a uniform incompatible strain and reduce the corresponding elastic energy by
\begin{equation}
 \Delta f_{\rm scr}
 =\frac{\mu_L^2e_0^2}{2(\mu_L+A_L)}.
 \label{eq:mechanismGain}
\end{equation}
The important point is that the matrix is neither simply ``soft'' nor simply ``solid''.  Its transient rigidity creates the elastic energy that makes incompatibility costly, while its remaining local mobility provides the channel that lowers that cost.

An ordered region can then be favorable because it gains structural free energy as $\phi$ increases, but its mismatch with the surrounding disordered matrix generates incompatible stress \cite{Alexander,Kroner1981}. The liquid-like matrix is valuable because its non-affine response is easier than in the ordered core, $A_L\ll A_X$, so it acts as a stress-accommodating medium.  In the local structural landscape, the bare function $f_0(\phi)$ favors crystallization on the low-temperature side of the liquid--crystal transition, whereas the stress-relaxation contribution preferentially lowers liquid-like configurations because they have the smaller non-affine stiffness.  The G state is therefore produced by a local competition between structural ordering and stress accommodation.  The dressed core minimum determines the structural healing scale $\xi_G$, while the surrounding stress field is screened over
\begin{equation}
 \lambda_\sigma
 =\ell_{\rm NA}\sqrt{\mathcal S_L},
 \qquad
 \ell_{\rm NA}^2=\frac{D_L}{\mu_L}.
 \label{eq:mechanismLambda}
\end{equation}
The G phase is the full heterogeneous state composed of these ordered cores plus the continuous liquid-like matrix; the matrix by itself should not be identified with a separate phase.

At the opposite kinetic extreme, $\tau_{\rm obs}\lesssim\tau_p$ while $\tau_\alpha^{\rm bulk}\gg\tau_{\rm obs}$, the local non-affine screening mode cannot equilibrate.  The rapidly quenched material then arrests before the heterogeneous stress-accommodation texture develops and forms a comparatively homogeneous \emph{L-like glass}.  This is distinct from the G phase, which is also glassy but heterogeneous and more rigid.

Chemistry changes the same balance.  A larger $\mu_L$ at fixed $A_L$ increases the accommodation fraction $\mathcal S_L$ and the elastic energy available to be released by screening.  However, chemistry also changes $A_L$, $D_L$, $\ell_{\rm NA}$, the incompatibility amplitude $e_0$, the dressed structural scale $\xi_G$, and the kinetic times.  The shell thickness and core size therefore cannot be predicted from $\mu_L$ alone.  A useful material combination remains
\begin{equation}
 \kappa_G
 =\frac{\ell_{\rm NA}}{\xi_G}
 \sqrt{\frac{\mu_L}{\mu_L+A_L}},
 \label{eq:mechanismKappa}
\end{equation}
together with the stress-relaxation energy scale and the kinetic inequalities above.

A compact kinetic summary is
\begin{equation}
 \begin{array}{rcl}
 \tau_\alpha^{\rm bulk}\lesssim\tau_{\rm obs},
 \ \tau_X\lesssim\tau_{\rm obs}
 &\Longrightarrow&
 \text{global relaxation and crystal nucleation/growth},\\[3pt]
 \tau_p\lesssim\tau_{\rm obs}\lesssim\tau_\alpha^{\rm bulk},
 \ \tau_X\gtrsim\tau_{\rm obs}
 &\Longrightarrow&
 \text{viscoelastic screening and G texture},\\[3pt]
 \tau_{\rm obs}\lesssim\tau_p,
 \ \tau_\alpha^{\rm bulk}\gg\tau_{\rm obs}
 &\Longrightarrow&
 \text{local arrest and homogeneous L-like glass}.
 \end{array}
 \label{eq:threeRegimes}
\end{equation}
These inequalities are kinetic organizing principles, not universal state boundaries; nucleation barriers, composition, and the magnitude and statistics of the incompatibility field remain essential.

\subsection{Finite cores and inhibited coalescence}
\label{sec:finitecores}

It is important to distinguish the structural healing length from an independently imposed particle radius.  The quantity
\begin{equation}
 \xi_G=\sqrt{\frac{K_\phi}{f_G''(\phi_G)}}
\end{equation}
is the distance over which the order parameter can change appreciably between the ordered core and the surrounding liquid-like material.  It therefore sets the natural minimum interfacial/core scale, but the GL gradient term alone does \emph{not} forbid an ordered domain with radius $R\gg\xi_G$.  If the bulk crystal were the only energetic winner and incompatibility could be relaxed without cost, the ordered domains would continue to grow and coarsen.

Finite G cores require the additional elastic frustration that is central to the present theory.  Growing an ordered core by moving its boundary outward converts part of the compliant, non-affinely mobile liquid-like shell into the more rigid ordered material.  Since $A_X\gg A_L$, that newly ordered material has less ability to accommodate the frozen incompatible strain.  The structural free-energy gain obtained by increasing the core volume must therefore be compared with an increasing residual elastic cost.  A useful schematic free energy for one approximately spherical core is
\begin{equation}
 F_{\rm core}(R)
 =-\frac{4\pi}{3}R^3\Delta g_{\rm str}
 +4\pi R^2\gamma_{LX}
 +F_{\rm inc}(R;h_\sigma,\mu_L,A_L,\lambda_\sigma),
 \label{eq:FcoreR}
\end{equation}
where $\Delta g_{\rm str}>0$ is the structural bulk driving force toward ordering and $F_{\rm inc}$ is the residual incompatibility energy after the liquid-like surroundings have relaxed as much as the time scale permits.  The first term favors growth, the second is the interfacial cost, and the last term penalizes coherent growth when the ordered region cannot accommodate the incompatibility as efficiently as the liquid-like shell.  Below we show that, if the ordered core carries an approximately size-independent uncompensated coarse-grained Nye density, the accumulated incompatible strain grows as $e_{\rm inc}\sim\alpha_{\rm eff}R$, which gives the small-core scaling $F_{\rm inc}\propto R^5$.

The equilibrium core size, when one exists, is determined by
\begin{equation}
 \left.\frac{\dd F_{\rm core}}{\dd R}\right|_{R=R_*}=0,
 \qquad
 \left.\frac{\dd^2 F_{\rm core}}{\dd R^2}\right|_{R=R_*}>0.
 \label{eq:Rstar}
\end{equation}
Near the G-state crossover there is only a modest structural gain in making the ordered region much larger than the distance over which $\phi$ heals, while the loss of stress-accommodating liquid and the incompatibility penalty grow rapidly.  It is therefore natural for $R_*$ to be of order $\xi_G$.  We emphasize, however, that $R_*\sim\xi_G$ is an energetic consequence of the coupled structural--elastic problem, not a mathematical identity: $\xi_G$ is a healing length, not a hard geometrical cutoff.

This distinction is also useful in the context of classical nucleation theory.  In the standard treatment reviewed by Kelton and Greer \cite{KeltonGreer2010}, the critical nucleus radius is selected by the competition between bulk free-energy driving and interfacial cost.  It is therefore conceptually distinct from an interfacial correlation or healing length.  In the present theory, $\xi_G$ plays the latter role: it sets the characteristic width over which structural order heals between the G core and the surrounding liquid-like matrix.  The barrier radius $R_b$ and the stable radius $R_*$, instead, follow from the full core free energy, including the incompatibility contribution.  The fact that they are of order $\xi_G$ near the G crossover is thus a physical consequence of the coupled problem, rather than an identification of the nucleation radius with the coherence length.

A second stabilization mechanism appears once more than one core is present.  The non-affine shell screens most of the incompatibility-generated stress but, for finite $A_L$ and finite response time, the screening is generally incomplete.  Each core therefore carries a residual deviatoric stress field.  In the one-channel approximation its far-field form is of screened-elastic type,
\begin{equation}
 \sigma^{(a)}(r)\sim \sigma_a^{\rm res}\,\frac{R_a}{r}
 \exp\!\left[-\frac{r-R_a}{\lambda_\sigma}\right].
 \label{eq:residualTail}
\end{equation}
When two such fields overlap, the quadratic elastic energy contains the cross term
\begin{equation}
 U_{12}
 =\int \dd^3r\,
 \sigma_{ij}^{(1)}(\bm r)\,
 S_{ijkl}\,
 \sigma_{kl}^{(2)}(\bm r),
 \label{eq:stressCross}
\end{equation}
where $S_{ijkl}$ is the appropriate elastic compliance tensor of the screening medium.  For two cores carrying the same projected deviatoric/incompatibility channel---meaning, in the one-channel reduction, that their residual scalar amplitudes are referred to the same tensor eigenmode $Q_{ij}$ and have the same sign, $\sigma_1^{\rm res}\sigma_2^{\rm res}>0$---this cross term is positive.  At separations larger than the core radii its leading form is therefore schematically Yukawa-like,
\begin{equation}
 U_{12}(d)\simeq U_0\,\frac{\exp(-d/\lambda_\sigma)}{d},
 \qquad U_0>0,
 \label{eq:YukawaCore}
\end{equation}
so that same-sign residual fields in the chosen projected channel repel once their stress-accommodation clouds begin to overlap.  This repulsion competes with the reduction in structural interfacial area that would otherwise favor coalescence.  The physical statement is therefore not that ``liquid shells repel'' in themselves; rather, overlap of \emph{like residual incompatible stress fields} costs elastic energy.  Because incompatibility is tensorial, the sign can depend on relative orientation: opposite or differently oriented channels can in principle attract.  The repulsive result in Eq.~\eqref{eq:YukawaCore} applies only to this same-sign projected-channel situation, which is the closest structural analogue of same-sign vortices.

The combined picture is consequently one of frustrated coarsening.  Structural ordering drives the formation and enlargement of ordered regions, while incompatibility makes unlimited coherent growth costly; the liquid-like shell lowers that cost locally through non-affine accommodation, and the residual screened fields provide an additional core--core interaction that can oppose coalescence.  In the G window this competition can stabilize a population of finite ordered cores, with a characteristic core scale of order $\xi_G$ and a characteristic interaction/screening range of order $\lambda_\sigma$.

\section{Finite-core stabilization and the structural mixed-state wedge}
\label{sec:numericalillustration}

The theory makes two nontrivial structural predictions that can be illustrated within one internally consistent parameter set: incompatibility can arrest coarsening at a finite nanometric core radius, and the same competition can open a stress--temperature G-state wedge between crystal and liquid.  The calculation below is deliberately coarse grained and is not a quantitative fit to Ag.  Its purpose is to demonstrate that the proposed mechanism is energetically self-consistent with physically reasonable mechanical and interfacial scales and to identify combinations of parameters that can be measured directly in simulation.

We deliberately use the term \emph{state diagram} rather than \emph{phase diagram}.  Although the GL and Gibbs-like constructions are equilibrium free-energy descriptions, the experimentally or computationally realized L, G, and X states are not necessarily mutually equilibrated on the available time scale.  Their accessibility also depends on the kinetic hierarchy among $\tau_{\rm obs}$, the bulk structural relaxation time, the local non-affine response time, and the crystal nucleation/growth time.  The boundaries drawn below should therefore be interpreted as boundaries between favored coarse-grained states under specified conditions, not as fully reversible equilibrium phase boundaries.

\subsection{A finite core with radius of order 1.5--2 nm}

We first choose a representative transient liquid shear modulus and non-affine stiffness,
\begin{equation}
 \mu_L=12~\mathrm{GPa},
 \qquad
 A_L=3~\mathrm{GPa}.
 \label{eq:numMuA}
\end{equation}
The corresponding accommodation fraction is
\begin{equation}
 \mathcal S_L=\frac{\mu_L}{\mu_L+A_L}=0.80.
\end{equation}
For an illustrative projected incompatible deviatoric strain $e_0=0.12$, the bare stress and the residual stress after local non-affine accommodation are
\begin{equation}
 h_\sigma=\mu_Le_0=1.44~\mathrm{GPa},
 \qquad
 \sigma_{\rm res}=h_\sigma\frac{A_L}{\mu_L+A_L}=0.288~\mathrm{GPa}.
 \label{eq:numResidualStress}
\end{equation}
The associated local energy density that can be released by non-affine stress accommodation is
\begin{equation}
 \Delta f_{\rm scr}
 =\frac{\mu_L^2e_0^2}{2(\mu_L+A_L)}
 =6.91\times10^7~\mathrm{J\,m^{-3}}.
 \label{eq:numDfscr}
\end{equation}
We take a representative stress-screening length
\begin{equation}
 \lambda_\sigma=3.0~\mathrm{nm},
\end{equation}
which corresponds to
\begin{equation}
 D_L=\lambda_\sigma^2(\mu_L+A_L)
 =1.35\times10^{-7}~\mathrm{J\,m^{-1}},
 \qquad
 \ell_{\rm NA}=\sqrt{\frac{D_L}{\mu_L}}=3.35~\mathrm{nm}.
 \label{eq:numDL}
\end{equation}

To illustrate how incompatibility can stop the growth of an ordered inclusion, we use the spherical-core free energy
\begin{equation}
 F_{\rm core}(R)
 =4\pi\left[
 \gamma R^2
 -\frac{\Delta g_{\rm loc}}{3}R^3
 +\frac{B_{\rm fr}}{5}\frac{R^5}{\lambda_\sigma^2}
 \right].
 \label{eq:FcoreIllustrative}
\end{equation}
The three terms have distinct physical roles.  The $R^2$ term is the liquid--ordered interfacial cost and creates the small-radius nucleation barrier.  The negative $R^3$ term is the net local free-energy drive toward formation of the ordered core.  The $R^5$ term represents the growing elastic cost of maintaining a coherently incompatible ordered region.  Its scaling can be motivated directly from the Nye/dislocation-density field defined in the Appendix, Eq.~\eqref{eq:nye}, rather than introduced only as an abstract super-extensive penalty.

Consider a coarse-grained projected Nye density $\alpha_{\rm eff}$ associated with the incompatibility carried by the ordered core.  The flux of the Nye tensor through a surface gives the corresponding Burgers-vector content, schematically
\begin{equation}
 b_{\rm eff}(S)\sim\int_S \alpha_{ai}n_i\,\dd S.
 \label{eq:beffAlpha}
\end{equation}
If the uncompensated projected density $\alpha_{\rm eff}$ is approximately independent of the core size, the Burgers content through a cross section of a spherical core scales as
\begin{equation}
 b_{\rm eff}\sim \alpha_{\rm eff}R^2.
 \label{eq:beffScaling}
\end{equation}
The same result follows locally from $\alpha\sim\nabla\times\beta^e$: for a distortion varying on the scale $R$, one has $\alpha_{\rm eff}\sim e_{\rm inc}/R$.  The incompatible strain accumulated across the core therefore scales as
\begin{equation}
 e_{\rm inc}\sim\frac{b_{\rm eff}}{R}\sim\alpha_{\rm eff}R.
 \label{eq:eincAlpha}
\end{equation}
Since elastic energy is quadratic in strain, the residual incompatibility energy density scales as $f_{\rm inc}\sim\mu_{\rm eff}e_{\rm inc}^2$, while the volume of the core scales as $R^3$.  Consequently
\begin{equation}
 F_{\rm inc}(R)\sim\mu_{\rm eff}\alpha_{\rm eff}^2R^5.
 \label{eq:FincR5alpha}
\end{equation}
Thus the $R^5$ behavior has a simple geometrical origin: a fixed coarse-grained incompatibility density gives an accumulated incompatible strain proportional to $R$, whose squared elastic cost is then integrated over a volume proportional to $R^3$.  Introducing the stress-screening scale through $\alpha_{\rm eff}\sim e_*/\lambda_\sigma$ gives
\begin{equation}
 F_{\rm inc}(R)\sim\mu_{\rm eff}e_*^2\frac{R^5}{\lambda_\sigma^2},
 \label{eq:FincR5lambda}
\end{equation}
which is the scaling used in Eq.~\eqref{eq:FcoreIllustrative}.  The coefficient $B_{\rm fr}$ therefore collects the effective modulus, the squared amplitude of the uncompensated projected incompatibility, and geometry/tensor projection factors; parametrically, $B_{\rm fr}\sim\mu_{\rm eff}(\alpha_{\rm eff}\lambda_\sigma)^2$.  The $R^5$ law is not asserted to be universal.  It relies on the physically explicit assumption that the core incorporates an approximately size-independent \emph{uncompensated coarse-grained} Nye density.  Strong cancellation between oppositely oriented defect content, or substantial redistribution of the incompatibility as the core grows, can modify this scaling and should ultimately be tested directly in MD.

For the numerical calibration we set $\Delta g_{\rm loc}=\Delta f_{\rm scr}$ as an energy-scale benchmark.  This identification is only illustrative: in a quantitative treatment $\Delta g_{\rm loc}$ contains the structural driving force and the stress-relaxation contribution separately.

\begin{figure}[t]
 \centering
 \includegraphics[width=0.78\linewidth]{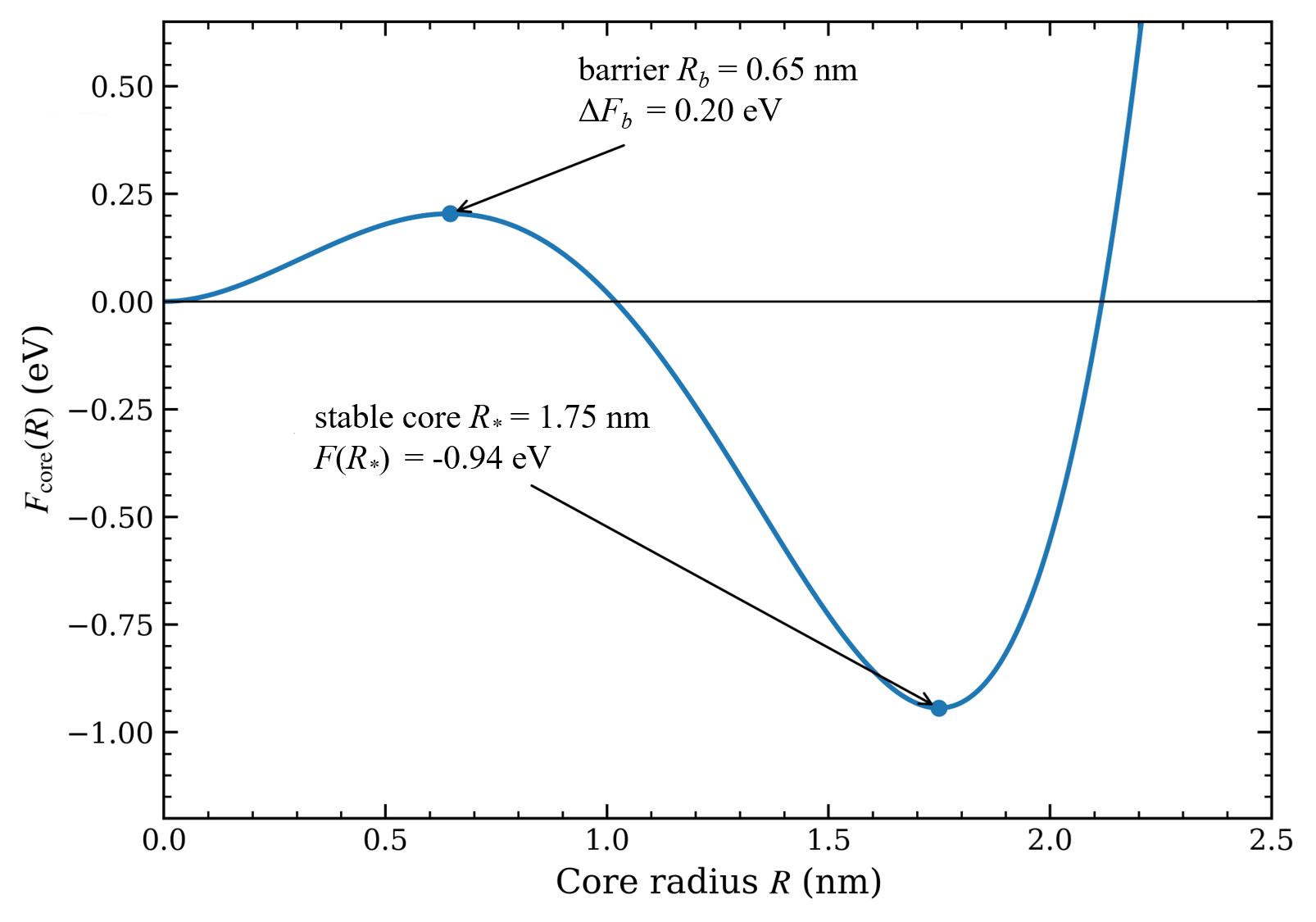}
 \caption{Illustrative finite-core free energy from Eq.~\eqref{eq:FcoreIllustrative} (blue curve).  The vertical scale is deliberately restricted to make the small nucleation barrier visible.  With the parameter set in Eqs.~\eqref{eq:numMuA}--\eqref{eq:numGammaB}, the barrier occurs at $R_b\simeq0.64$ nm, while incompatibility frustration generates a stable minimum at $R_*\simeq1.75$ nm.  The numerical values are an internally consistent demonstration rather than a fit to a particular alloy.}
 \label{fig:corecalibration}
\end{figure}

Choosing
\begin{equation}
 \gamma=20.3~\mathrm{mJ\,m^{-2}},
 \qquad
 B_{\rm fr}=1.35\times10^8~\mathrm{J\,m^{-3}},
 \label{eq:numGammaB}
\end{equation}
leads to the stationarity condition
\begin{equation}
 \frac{1}{4\pi R}\frac{\dd F_{\rm core}}{\dd R}
 =2\gamma-\Delta g_{\rm loc}R
 +B_{\rm fr}\frac{R^3}{\lambda_\sigma^2}=0.
 \label{eq:numStationarity}
\end{equation}
There are two positive stationary radii.  The first,
\begin{equation}
 R_b=0.64~\mathrm{nm},
 \qquad
 \Delta F_b=0.20~\mathrm{eV},
\end{equation}
is a nucleation barrier.  The second,
\begin{equation}
 R_*=1.75~\mathrm{nm},
 \qquad
 F_{\rm core}(R_*)=-0.95~\mathrm{eV},
 \label{eq:numRstar}
\end{equation}
is a stable finite-radius minimum.  Figure~\ref{fig:corecalibration} shows the result on an energy scale chosen so that the small nucleation barrier remains visible.  The calculation demonstrates that the coexistence of a modest interfacial cost, a local ordering/relaxation drive, and a growing incompatibility penalty is sufficient to stabilize cores in the $1.5$--$2$ nm range without imposing that radius by hand.  The radius $R_*$ should still be distinguished conceptually from the healing length $\xi_G$; in the G regime the two are expected to be of the same order, but $R_*$ is selected by the full core free energy whereas $\xi_G$ measures the width over which structural order heals.

\begin{table}[t]
 \caption{Illustrative parameter set used for the finite-core calculation.  The mechanical values are intended to be representative of a deeply undercooled viscoelastic liquid on the G-formation time scale, not equilibrium zero-frequency liquid constants.}
 \label{tab:illustrativeparameters}
 \begin{ruledtabular}
 \begin{tabular}{lc}
 Quantity & Illustrative value \\
 \hline
 $\mu_L$ & $12$ GPa \\
 $A_L$ & $3$ GPa \\
 $\mathcal S_L=\mu_L/(\mu_L+A_L)$ & $0.80$ \\
 $e_0$ & $0.12$ \\
 $h_\sigma$ & $1.44$ GPa \\
 $\sigma_{\rm res}$ & $0.288$ GPa \\
 $\lambda_\sigma$ & $3.0$ nm \\
 $\ell_{\rm NA}$ & $3.35$ nm \\
 $\gamma$ & $20.3$ mJ m$^{-2}$ \\
 $B_{\rm fr}$ & $1.35\times10^8$ J m$^{-3}$ \\
 \end{tabular}
 \end{ruledtabular}
\end{table}

\subsection{Stress as the structural analogue of magnetic field}

To construct the analogue of the superconducting $H$--$T$ phase diagram, we introduce a stress--temperature state diagram based on a Gibbs-like potential at fixed projected deviatoric stress $\sigma$.  In a homogeneous phase $i=L,X$, the mechanical contribution is
\begin{equation}
 g_i(T,\sigma)=g_i^0(T)-\frac{\sigma^2}{2m_i},
 \qquad
 m_i>0,
 \label{eq:giStress}
\end{equation}
where $m_i$ is the effective stiffness of that phase in the chosen mechanically allowed deviatoric channel.  The liquid-like phase is taken to be more compliant on the relevant time scale, $m_L<m_X$, so increasing stress lowers the liquid free energy more strongly than the crystal free energy.  The stress axis in what follows should therefore be read as an \emph{effective projected incompatible stress amplitude}.  In the real G problem this need not be a uniform externally applied stress; it may instead represent an rms or core-scale coarse-grained amplitude of the internal, mechanically equilibrated deviatoric stress.

For an Ag-like reference melting temperature we choose
\begin{equation}
 T_m=1235~\mathrm{K},
 \qquad
 L_v=1.10\times10^9~\mathrm{J\,m^{-3}},
\end{equation}
where $L_v$ is the latent heat (enthalpy of fusion) per unit volume.  Near $T_m$, taking the entropy-of-fusion density as approximately $L_v/T_m$ gives the simple bulk liquid--crystal free-energy difference
\begin{equation}
 \Delta g_0(T)
 \equiv g_L^0-g_X^0
 =L_v\left(1-\frac{T}{T_m}\right)
 \qquad (T<T_m).
 \label{eq:DeltaG0T}
\end{equation}
With
\begin{equation}
 m_L=\mu_L+A_L=15~\mathrm{GPa},
 \qquad
 m_X=60~\mathrm{GPa},
\end{equation}
the stress-dressed liquid--crystal difference becomes
\begin{equation}
 \Delta g_{LX}(T,\sigma)
 =L_v\left(1-\frac{T}{T_m}\right)
 -C_\sigma\sigma^2,
 \label{eq:DeltaGLXStress}
\end{equation}
with
\begin{equation}
 C_\sigma
 =\frac12\left(\frac{1}{m_L}-\frac{1}{m_X}\right)
 =2.50\times10^{-11}~\mathrm{Pa^{-1}}.
\end{equation}
If no heterogeneous G state were available, the direct liquid--crystal coexistence line would be
\begin{equation}
 \sigma_c(T)
 =\sqrt{\frac{L_v(1-T/T_m)}{C_\sigma}}.
 \label{eq:sigmaDirect}
\end{equation}
This is the structural analogue of the single thermodynamic critical field of a type-I superconductor.

To represent the energetic advantage of the heterogeneous state in the simplest possible manner, let
\begin{equation}
 g_G(T,\sigma)
 =\frac{g_L(T,\sigma)+g_X(T,\sigma)}{2}
 -\epsilon_G(T),
 \label{eq:gGphenom}
\end{equation}
where $\epsilon_G>0$ is the coarse-grained stabilization energy density obtained by combining ordered-core formation with stress accommodation in the liquid-like material.  For the illustrative diagram we take
\begin{equation}
 \epsilon_G(T)=\epsilon_0\left(1-\frac{T}{T_m}\right),
 \qquad
 \epsilon_0=0.07L_v.
 \label{eq:epsilonGmodel}
\end{equation}
This term is phenomenological and should ultimately be obtained from the core density, interfacial energy, and incompatibility-relaxation energy measured in MD.  Its purpose here is only to quantify how an intermediate G wedge opens once the heterogeneous texture has a finite stabilization energy.

The crystal--G boundary follows from $g_X=g_G$, or $\Delta g_{LX}=2\epsilon_G$, while the G--liquid boundary follows from $g_G=g_L$, or $\Delta g_{LX}=-2\epsilon_G$.  The two critical stresses are therefore
\begin{equation}
 \sigma_{c1}(T)
 =\sqrt{\frac{[L_v-2\epsilon_0](1-T/T_m)}{C_\sigma}},
 \label{eq:sigmac1}
\end{equation}
\begin{equation}
 \sigma_{c2}(T)
 =\sqrt{\frac{[L_v+2\epsilon_0](1-T/T_m)}{C_\sigma}}.
 \label{eq:sigmac2}
\end{equation}
The resulting ordering is
\begin{equation}
 \begin{array}{lll}
 \sigma<\sigma_{c1}(T) &\Longrightarrow& X\ \text{(crystal)},\\[3pt]
 \sigma_{c1}(T)<\sigma<\sigma_{c2}(T) &\Longrightarrow& G,\\[3pt]
 \sigma>\sigma_{c2}(T) &\Longrightarrow& L\ \text{(liquid)}.
 \end{array}
 \label{eq:stressRegimes}
\end{equation}
The direct line $\sigma_c(T)$ lies inside the G wedge.  If the heterogeneous stabilization disappears, $\epsilon_G\rightarrow0$, Eqs.~\eqref{eq:sigmac1} and \eqref{eq:sigmac2} collapse onto Eq.~\eqref{eq:sigmaDirect}; this is the structural type-I-like limit.  A finite $\epsilon_G$ separates the two boundaries and produces the type-II-like sequence crystal $\rightarrow$ G $\rightarrow$ liquid.

\begin{figure}[t]
 \centering
\includegraphics[width=0.78\linewidth]{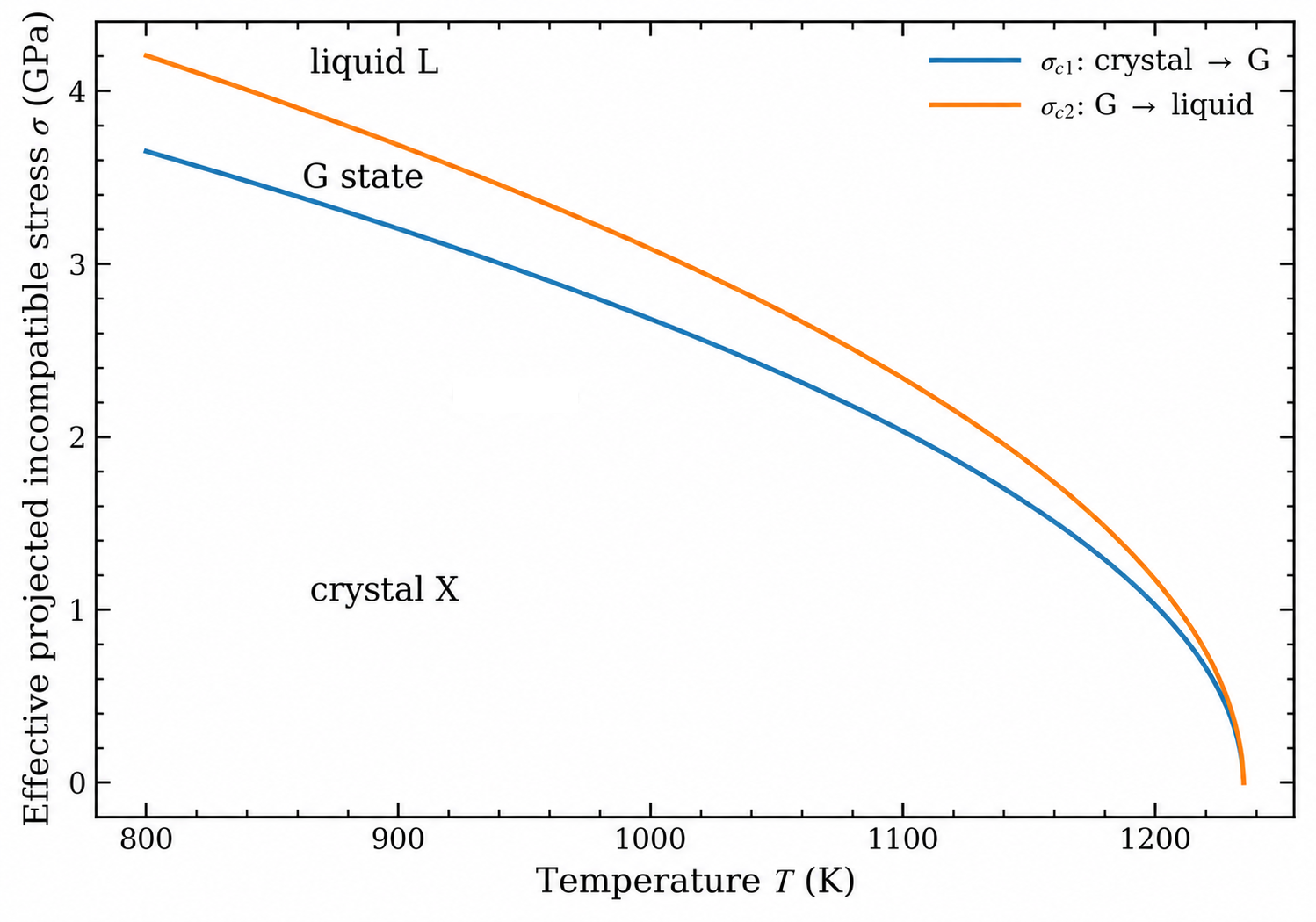}
 \caption{Illustrative temperature--stress state diagram showing the two physical boundaries $\sigma_{c1}$ and $\sigma_{c2}$ from Eqs.~\eqref{eq:sigmac1} and \eqref{eq:sigmac2}.  The vertical axis is the effective projected incompatible deviatoric stress, playing the organizing role of the magnetic field in the superconducting analogy.  At low stress the crystal is favored; at intermediate stress the heterogeneous G state is favored; at sufficiently large stress the liquid wins because its lower effective stiffness allows a larger mechanical free-energy reduction.  The direct L--X coexistence line of Eq.~\eqref{eq:sigmaDirect} lies inside the G wedge and is omitted from the figure for clarity; it is recovered when the heterogeneous stabilization $\epsilon_G$ vanishes.  The diagram is schematic and uses illustrative parameters rather than a fit to Ag.}
 \label{fig:stressphasediagram}
\end{figure}

For example, at $T=1000$ K the illustrative parameters give
\begin{equation}
 \sigma_{c1}=2.68~\mathrm{GPa},
 \qquad
 \sigma_c=2.89~\mathrm{GPa},
 \qquad
 \sigma_{c2}=3.09~\mathrm{GPa},
\end{equation}
while at $T=900$ K they give approximately $3.20$, $3.46$, and $3.69$ GPa, respectively.  Figure~\ref{fig:stressphasediagram} shows the complete illustrative state diagram.

The numerical exercise clarifies the meaning of the inverted Abrikosov analogy.  The lower boundary $\sigma_{c1}$ is the stress at which maintaining finite ordered cores plus stress-accommodating liquid first becomes preferable to a homogeneous crystal.  The upper boundary $\sigma_{c2}$ is the stress at which even the ordered cores cease to be advantageous and the liquid becomes the lowest-free-energy state.  Between them, the system gains from having both components simultaneously: ordered regions exploit the structural drive toward crystallinity, while the surrounding viscoelastic liquid lowers the incompatibility energy.  The finite value of $R_*$ is then set by the competition between interfacial energy, local free-energy gain, and the increasing incompatibility penalty, while $\lambda_\sigma$ controls the spatial range over which the surrounding liquid can accommodate and screen the residual stress.

An important kinetic asymmetry should be emphasized.  The state diagram is not meant to imply that all passages between L, G, and X are reversible on laboratory time scales.  A G state can be rejuvenated toward an L-like state, while an initially crystalline X state can in principle be driven into a G-like disordered heterogeneous state by processes that create defects and incompatibility, such as ion bombardment, irradiation, or severe localized shear/shear-band formation.  By contrast, a spontaneous G$\rightarrow$X conversion by ordinary ageing is expected to be effectively inaccessible on experimentally relevant time scales.  In the corresponding energy-landscape picture, the metastable G minimum lies only modestly above the X minimum, so the residual thermodynamic driving force for further ordering is small and the associated nucleation-and-growth time can become prohibitively long.  This differs from conventional glass$\rightarrow$crystal ageing when the glass occupies a substantially higher-energy minimum relative to a much deeper crystalline minimum and crystallization can therefore be observed.  The arrows connecting the states are consequently history- and protocol-dependent, which is another reason to regard Fig.~\ref{fig:stressphasediagram} as a state diagram rather than a fully equilibrated phase diagram.

\section{Direct observables and falsifiable tests}

A central requirement for a broad mixed-state interpretation is that the theory be testable through observables rather than only through fitted free-energy coefficients.  The transient modulus $\mu_L$ should be obtained from the shear-stress relaxation function or from a small-amplitude oscillatory/step-shear protocol on the observation/transformation time scale, so that $\mu_L\simeq G'(\omega_{\rm obs},T)$ or from the corresponding plateau modulus.  Measuring $G''$ at the same frequency locates the viscoelastic crossover, while the stored-energy coefficient entering the free energy is associated with the recoverable part $G'$.  The coefficient $A_L$ can then be separated from $\mu_L$ through the non-affine susceptibility, and $D_L$ from its wave-vector dependence.

For the homogeneous liquid channel, a small projected source obeys in Fourier space
\begin{equation}
 p(k)=\frac{h_\sigma(k)}{\mu_L+A_L+D_Lk^2}.
\end{equation}
Thus
\begin{equation}
 \chi_p^{-1}(k)=\mu_L+A_L+D_Lk^2.
 \label{eq:chipMD}
\end{equation}
The intercept determines $\mu_L+A_L$ and the slope determines $D_L$; an independent measurement of $\mu_L$ then gives $A_L$.  The same data give
\begin{equation}
 \ell_{\rm NA}=\sqrt{\frac{D_L}{\mu_L}},
 \qquad
 \lambda_\sigma=\sqrt{\frac{D_L}{\mu_L+A_L}}.
\end{equation}
Alternatively, $\lambda_\sigma$ can be obtained directly from the spatial decay of the mechanically projected deviatoric stress around a G core or interface.

The local accommodation fraction can also be measured directly by comparing the projected incompatible strain before and after non-affine relaxation,
\begin{equation}
 \mathcal S_L=\frac{p_0}{e_0}
 =\frac{\mu_L}{\mu_L+A_L},
\end{equation}
while the residual-stress fraction is $1-\mathcal S_L$.  The temperature and cooling-rate dependence of these quantities is as important as their absolute values.  In particular, the static susceptibility should be supplemented by the frequency-dependent response in Eq.~\eqref{eq:pdynamicSusceptibility} to determine whether the $p$ mode is dynamically accessible during G-phase formation.

For the actual G state, the long-wavelength structural fluctuation spectrum should be interpreted in terms of the dressed curvature,
\begin{equation}
 S_\phi^{-1}(k)\propto f_G''(\phi_G)+K_\phi k^2,
 \label{eq:SphiMD}
\end{equation}
so that the intercept-to-slope ratio gives the physical $\xi_G^2=K_\phi/f_G''(\phi_G)$.  A corresponding stress-free or weak-stress calibration gives the bare reference $\xi_0^2=K_\phi/f_0''(\phi_X)$.  Comparing $\xi_G$ with $\xi_0$ is therefore a direct way to quantify how incompatibility and mechanical response reshape the structural landscape.  A second route to $K_\phi$ follows from the bare structural interfacial energy and is derived below.

The decisive dynamical test is independent of these static extractions.
The frequency dependence of the non-affine susceptibility determines the
local response time $\tau_p$, which should be compared with the protocol
time $\tau_{\rm obs}$, the bulk structural relaxation time
$\tau_\alpha^{\rm bulk}$, and the crystal nucleation-and-growth time
$\tau_X$.  The theory predicts preferential G formation when
$\tau_p\lesssim\tau_{\rm obs}\lesssim\tau_\alpha^{\rm bulk}$ while
$\tau_X\gtrsim\tau_{\rm obs}$.  If a Johari--Goldstein $\beta$ process is
independently resolved, comparing its time scale with $\tau_p$ directly
tests whether that secondary relaxation participates in the
incompatibility-screening channel rather than assuming such an
identification.

A second stringent test concerns spatial interactions.  Stress profiles
conditioned on individual G cores should decay over $\lambda_\sigma$,
whereas the structural profile should heal over $\xi_G$.  For two nearby
cores, the constrained excess free energy or potential of mean force
should depend on both separation and relative incompatibility
orientation.  Same-sign residual amplitudes in a common projected
deviatoric channel are predicted to interact repulsively over
$\lambda_\sigma$, whereas other tensor orientations need not.  Measuring
these quantities would test the two ingredients that distinguish the
present mechanism from a purely phenomenological two-phase mixture:
finite mechanical screening and frustration of coarsening by residual
incompatibility.

\section{Structural interface scale and energetic stabilization of the G texture}

\subsection{Exact structural kink and the relation between \texorpdfstring{$K_\phi$}{Kphi} and interfacial energy}

At stress-free coexistence the sixth-order potential factorizes as
\begin{equation}
 f_0(\phi)=\frac{c}{6}\phi^2(\phi^2-\phi_X^2)^2.
\end{equation}
For a planar interface the first integral of the bare structural GL equation is
\begin{equation}
 \frac{K_\phi}{2}(\phi')^2=f_0(\phi).
 \label{eq:firstintegral}
\end{equation}
Choosing liquid at $x\to-\infty$ and ordered material at $x\to+\infty$ gives the exact profile
\begin{equation}
 \phi^2(x)=\frac{\phi_X^2}
 {1+\exp[-(x-x_0)/\xi_0]},
 \label{eq:exactkink2}
\end{equation}
with
\begin{equation}
 \xi_0=\sqrt{\frac{K_\phi}{f_0''(\phi_X)}}
 =\sqrt{\frac{4K_\phi c}{3b^2}}.
\end{equation}
The corresponding bare structural interface tension is
\begin{equation}
 \gamma_\phi
 =\int_{-\infty}^{\infty}\dd x
 \left[\frac{K_\phi}{2}(\phi')^2+f_0(\phi)\right]
 =\frac{K_\phi\phi_X^2}{8\xi_0}.
 \label{eq:gammaKxi}
\end{equation}
Hence
\begin{equation}
 K_\phi=\frac{8\gamma_\phi\xi_0}{\phi_X^2},
 \label{eq:Kfromgamma}
\end{equation}
and $K_\phi=8\gamma_\phi\xi_0$ if $\phi_X$ is normalized to unity.  This relation is useful because it connects the phenomenological gradient coefficient to two quantities that can be measured independently in simulation: a structural interface free energy and an interface width.  The numerical coefficient is specific to the chosen sixth-order potential and normalization.
The length in this analytic kink is the \emph{bare} reference $\xi_0$, because the calculation uses $f_0$ alone.  The actual G-core profile is mechanically dressed and need not have the same width; its healing length is $\xi_G=\sqrt{K_\phi/f_G''(\phi_G)}$.  Thus Eq.~\eqref{eq:Kfromgamma} is a calibration of the structural gradient coefficient from the stress-free reference interface, not a statement that incompatible stress leaves the G-core size unchanged.

\subsection{Coupling the structural interface to stress accommodation}

The simplest coupled one-dimensional description keeps the incompatibility-generated bare source $h_\sigma(x)$ explicit and allows the local non-affine stiffness to depend on structure,
\begin{equation}
 A(0)=A_L,
 \qquad
 A(\phi_X)=A_X,
 \qquad
 A_X\gg A_L.
\end{equation}
The coupled equations are then
\begin{equation}
 -K_\phi\phi''+f_0'(\phi)+\frac12A'(\phi)p^2=0,
 \label{eq:phiCoupled1d}
\end{equation}
\begin{equation}
 -D_Lp''+[\mu_L+A(\phi)]p=h_\sigma(x).
 \label{eq:pCoupled1d}
\end{equation}
These equations show directly how the two parts of the mechanism communicate.  The structural order controls how easily non-affine accommodation can occur, while the non-affine field changes the free-energy balance across the interface.  No additional ``effective'' local potential is required to state this coupling.

The full liquid--ordered interface tension is an excess free energy.  For a planar system of area $\mathcal A$ it is defined by
\begin{equation}
 \gamma_{LX}
 =\frac{F_{\rm int}-F_{\rm bulk}}{\mathcal A},
 \qquad
 F_{\rm bulk}=V_Lg_L^{\rm bulk}+V_Xg_X^{\rm bulk}.
 \label{eq:FbulkExplicit}
\end{equation}
At a stationary planar coexistence interface, $g_L^{\rm bulk}=g_X^{\rm bulk}$.  The subtraction is important because bulk stress-relaxation energy must not be counted as interfacial energy.

The type-II analogy should therefore be used as an energetic organizing principle rather than as a universal criterion.  A finite G texture competes with two limiting tendencies: reducing interface area favors coarsening toward the crystal, while retaining liquid-like material can be favorable because that material accommodates incompatible stress more efficiently than the ordered phase.  The balance depends on $\gamma_\phi$, the incompatibility amplitude, $\Delta f_{\rm scr}$, $\lambda_\sigma$, and the kinetics of the $p$ response.  We may define a material- and protocol-dependent crossover $\kappa_G^*$ by
\begin{equation}
 \gamma_{LX}(\kappa_G^*;\text{source, chemistry, rate})=0,
 \label{eq:kappastarGeneral}
\end{equation}
but there is no reason for $\kappa_G^*$ to equal the superconducting value $1/\sqrt2$, nor is $\kappa_G$ alone sufficient to determine the G phase.  For the present theory the most robust prediction is the physical competition summarized in Secs.~\ref{sec:mechanism} and \ref{sec:finitecores}: structural ordering favors the core, incompatibility penalizes uncontrolled coherent growth, a viscoelastic liquid shell lowers that penalty through local non-affine accommodation, and any residual screened stress produces an additional interaction between neighboring cores that can oppose coalescence.

\section{Conclusions}
\label{sec:conclusions}

We have developed a mechanically consistent Ginzburg--Landau framework
for the heterogeneous G state of deeply undercooled metallic liquids.
The central result is that an inverted type-II-like structural state can
emerge from ordinary ingredients of continuum mechanics once three
objects that are often conflated are kept distinct: elastic incompatibility between ordered cores and the surrounding liquid-like amorphous matrix is
the source, mechanically equilibrated stress is the response, and
non-affine rearrangement is the screening channel. The analogy with a
type-II superconductor  emerges \emph{mutatis mutandis} from the competition between a
stress-screening length (analogous to a London length) and a mechanically dressed structural healing
length.

This construction gives a unified physical explanation for why the
liquid-like matrix must be neither fully fluid nor fully frozen.  A
finite transient shear rigidity makes frozen incompatibility
energetically costly, while residual local mobility allows non-affine
rearrangements to lower that energy cost. The mixed G state (ordered cores in amorphous liquid-like matrix) is consequently selected
only over an intermediate dynamical window.  Longer available times
permit global relaxation and, when nucleation is accessible,
crystallization; shorter times arrest even the local non-affine stress-accommodation mode
and favor a comparatively homogeneous L-like glass. The same mechanism
thus links morphology and kinetics rather than treating the G texture
and the cooling-rate dependence as separate phenomena.

The mechanical coupling also feeds back on structural order.  Stress
relaxation dresses the local Landau landscape and therefore changes the
structural healing length $\xi_G$, while the non-affine field produces
the London-type screening length $\lambda_\sigma$.  Their ratio
$\kappa_G=\lambda_\sigma/\xi_G$ organizes the geometry of the mixed
G texture, although it is not by itself a universal stability criterion.
Finite core radii arise only after interfacial and incompatibility
energies are included.  Residual screened stresses can additionally
frustrate coalescence of the cores in an orientation-dependent manner. The resulting
state is thus an inverted Abrikosov analogue at the level of physical
roles: structural order is concentrated in the cores, while the
surrounding disordered matrix carries the dominant screening response.

The broader implication is that type-II-like mixed-state phenomenology
need not be tied to electromagnetism or to a literal gauge-field
mapping.  In a purely structural condensed matter problem, a comparable organization can arise
when an order parameter heals over one length scale while a mechanically
generated frustration field is relaxed over another. In the present
case, those fields are connected by incompatible elasticity and
non-affine screening, which makes the construction directly testable
rather than purely analogical. Stress-decay and structural-healing
profiles should independently determine $\lambda_\sigma$ and $\xi_G$;
the non-affine susceptibility should determine the local accommodation
and cooperativity parameters; and the kinetic hierarchy should predict
which of the crystal, G, or L-like glass states is selected by a given
protocol.

Several ingredients remain deliberately coarse grained, most notably the
statistics of the incompatibility source, the structural dependence of
the non-affine stiffness, and the detailed size dependence of the
residual elastic energy.  The illustrative $R^5$ incompatibility term is
therefore not claimed to be universal, and the stress--temperature
construction is a state diagram rather than a fully equilibrated phase
diagram.  These limitations are also opportunities for direct
simulation and experiment: measuring the two characteristic lengths,
the non-affine response time, and the orientation-dependent interaction
between neighboring cores provides a stringent route to deciding
whether incompatibility-driven screening is the mechanism underlying the
heterogeneous G state.

\appendix

\section{Elastic gauge formulation underlying the superconducting analogy}

A convenient superconducting Gibbs-type functional is \cite{Tinkham}
\begin{equation}
\mathcal G_{\rm SC}[\psi,\bm A;\bm H]
=
\int \dd^3r\,
\left[
 f_{\rm SC}(|\psi|)
 +K_s\left|\left(\bm\nabla-iq\bm A\right)\psi\right|^2
 +\frac{1}{2\mu_0}\left|\bm\nabla\times\bm A-\mu_0\bm H\right|^2
\right].
\label{eq:SCGL}
\end{equation}
The essential hierarchy is
\begin{equation}
\bm A\xrightarrow{\;\bm\nabla\times\;}\bm B,
\qquad
\bm B=\bm\nabla\times\bm A,
\label{eq:EMhierarchy}
\end{equation}
where $\bm A$ is a gauge potential, $\bm B$ is the induction, and $\bm H$ is the imposed control field. In the London limit $|\psi|\simeq\psi_0$, variation with respect to $\bm A$ produces a Helmholtz equation and hence exponential screening with penetration depth $\lambda$.

The expression first proposed for the structural problem was the direct transcription
\begin{equation}
F_{I}^{\rm naive}
=
\int \dd^3r\,
\left[
 f_0(|\Psi|)
 +K\left|\left(\bm\nabla-iq_s\bm A_s\right)\Psi\right|^2
 +\frac{1}{2\chi_s}
 \left|\bm\nabla\times\bm A_s-\bm h(\bm r)\right|^2
\right].
\label{eq:naive}
\end{equation}
Equation~\eqref{eq:naive} is useful as a mnemonic, but its original verbal interpretation must be corrected. In particular,
\begin{equation}
\bm\nabla\times\bm A_s\neq\text{stress}
\end{equation}
in a faithful elastic theory. The next section explains what the elastic gauge field can mean.

\section{Formulation I: elastic gauge field and incompatibility}

\subsection{Compatible distortion}

Let $u_a(\bm r)$ be the displacement field. A purely compatible distortion is
\begin{equation}
\beta_{ai}^{\rm c}=\partial_i u_a.
\label{eq:compatible_distortion}
\end{equation}
Because partial derivatives commute,
\begin{equation}
\epsilon_{ijk}\partial_j\beta_{ak}^{\rm c}=0.
\label{eq:curlcompatible}
\end{equation}
Thus a single-valued displacement has zero distortion curl.

\subsection{Incompatible distortion as a translational gauge field}

Introduce an incompatible distortion potential $\Ael_{ai}$ and define the elastic distortion
\begin{equation}
\beta^e_{ai}=\partial_i u_a-\Ael_{ai}.
\label{eq:elasticdistortion}
\end{equation}
Under the local translation-like transformation
\begin{equation}
u_a\rightarrow u_a+\zeta_a,
\qquad
\Ael_{ai}\rightarrow\Ael_{ai}+\partial_i\zeta_a,
\end{equation}
the combination \eqref{eq:elasticdistortion} is unchanged. This is the precise sense in which $\Ael_{ai}$ can be regarded as an elastic gauge potential in continuum defect theories \cite{Lazar2011final}.

Its field strength is the Nye/dislocation-density tensor
\begin{equation}
\alpha_{ai}
=\epsilon_{ijk}\partial_j\Ael_{ak}
=-\epsilon_{ijk}\partial_j\beta^e_{ak},
\label{eq:nye}
\end{equation}
up to sign convention. The electromagnetic analogy at this level is therefore
\begin{equation}
A_i^{\rm EM}\longleftrightarrow \Ael_{ai},
\qquad
B_i=\epsilon_{ijk}\partial_jA_k
\longleftrightarrow
\alpha_{ai}=\epsilon_{ijk}\partial_j\Ael_{ak}.
\label{eq:gauge_dictionary}
\end{equation}
The extra material index $a$ is unavoidable in full elasticity because translations have three components.

\subsection{From dislocation density to strain incompatibility}

The elastic strain is the symmetric part of the elastic distortion,
\begin{equation}
 e_{ij}=\frac{1}{2}\left(\beta^e_{ij}+\beta^e_{ji}\right).
\label{eq:strain}
\end{equation}
For a compatible strain the Saint--Venant condition is
\begin{equation}
\bm\nabla\times\bigl(\bm\nabla\times e\bigr)=0,
\end{equation}
where in components
\begin{equation}
\left[\bm\nabla\times\bigl(\bm\nabla\times e\bigr)\right]_{ij}
\equiv
\epsilon_{ikl}\epsilon_{jmn}
\partial_k\partial_m e_{ln}.
\label{eq:incdef}
\end{equation}
In the presence of defects we write
\begin{equation}
\left[\bm\nabla\times\bigl(\bm\nabla\times e\bigr)\right]_{ij}=\eta_{ij}.
\label{eq:incompatibility}
\end{equation}
The incompatibility tensor is related to gradients of the Nye tensor. With one common convention,
\begin{equation}
\eta_{ij}
=
-\frac{1}{2}
\left(
\epsilon_{ikl}\partial_k\alpha_{lj}
+
\epsilon_{jkl}\partial_k\alpha_{li}
\right).
\label{eq:kroner}
\end{equation}
This is the structural hierarchy that replaces the oversimplified identification $\nabla\times A_s\sim\sigma$:
\begin{equation}
\Ael
\xrightarrow{\Curl}
\alpha
\xrightarrow{\text{symmetrized Curl}}
\eta
\xrightarrow{\text{elastic Green function}}
\sigma.
\label{eq:structuralhierarchy}
\end{equation}
The incompatibility literature emphasizes precisely this point: defect-induced incompatibility generates long-range elastic contributions and stress fields in structural Landau theories \cite{Groger2008final,Groger2010final}.

\subsection{Stress is a response, not a gauge curvature}

The stress is obtained from the constitutive relation
\begin{equation}
\sigma_{ij}=C_{ijkl}e_{kl},
\label{eq:Hooke}
\end{equation}
and, in the absence of body force, must satisfy
\begin{equation}
\partial_j\sigma_{ij}=0.
\label{eq:mecheq}
\end{equation}
Equations~\eqref{eq:incompatibility}, \eqref{eq:Hooke}, and \eqref{eq:mecheq} determine the stress generated by a prescribed incompatibility. Symbolically,
\begin{equation}
\sigma_{ij}(\bm r)
=\int\dd^3r'\,
\mathcal G_{ijmn}(\bm r-\bm r')\,\eta_{mn}(\bm r'),
\label{eq:elasticGreen}
\end{equation}
where $\mathcal G$ is the elastic Green kernel, including the mechanical-equilibrium projection.

This immediately resolves the meaning of the random field in Eq.~\eqref{eq:naive}. The cleanest interpretation is not
\begin{equation}
\bm h=\text{stress},
\end{equation}
but rather
\begin{equation}
\bm h\ \text{in Eq.~\eqref{eq:naive}}
\;\leadsto\;
\text{a projected bare defect/incompatibility source}.
\label{eq:hreinterpret}
\end{equation}
If one wishes to compare directly with a measured or simulated local stress, define instead the \emph{bare stress field}
\begin{equation}
 h_{\sigma,ij}(\bm r)
 =\int\dd^3r'\,
\mathcal G_{ijmn}(\bm r-\bm r')\,\eta^{0}_{mn}(\bm r'),
\label{eq:barestress}
\end{equation}
where $\eta^0$ is the frozen or slowly evolving incompatibility before local non-affine relaxation.

\subsection{What remains useful from the superconducting-looking functional}

After projection onto a single dominant deviatoric channel one may still write a schematic gauge functional
\begin{equation}
F_I=
\int\dd^3r\left[
 f_0(|\Psi|)
 +K\left|\left(\nabla-iq_s a\right)\Psi\right|^2
 +\frac{1}{2\chi_\alpha}\left|\nabla\times a-h_\alpha\right|^2
\right],
\label{eq:FormI}
\end{equation}
provided that $a$ is understood as a projected incompatible distortion potential and $h_\alpha$ as a projected defect/incompatibility source, not as the physical stress. Equation~\eqref{eq:FormI} preserves the conceptual analogy with type-II superconductivity, but it is not yet the most transparent starting point for calculating the stress profile of a G-phase core.

\section{Why Formulation I should be reduced rather than taken literally}

There are three reasons to pass to a direct incompatible-elastic formulation.

First, the measured quantity of interest is stress or deviatoric strain, while the gauge curvature $\alpha$ is a defect-density object. Second, mechanical equilibrium, Eq.~\eqref{eq:mecheq}, is not automatically guaranteed by a scalar $|\nabla\times a-h|^2$ term. Third, ordinary incompatible elasticity is long ranged; it does not by itself generate a London penetration depth. A finite observed stress-screening length must therefore come from an additional relaxational degree of freedom of the liquid-like shell.

This motivates the main incompatible-elastic formulation used in the text: keep incompatibility explicit, but work directly with the physical fields needed to describe structural order and stress relaxation rather than continuing with the gauge notation.

\begin{acknowledgments}
A.Z. gratefully acknowledges the Nieders\"achsische Akademie der
Wissenschaften zu G\"ottingen for its support and hospitality in the frame of the Gauss Professorship. K.S. is thankful for many discussions with W. L. Johnson (Caltech), and with A. Loidl, P. Lunkenheimer and U. Eckern (University of Augsburg).
\end{acknowledgments}

\bibliographystyle{apsrev4-2}
\bibliography{references_GL}

\end{document}